\documentclass[twocolumn,showkeys,showpacs]{revtex4-1}
\usepackage[cp1251]{inputenc}
\usepackage{amsmath}
\usepackage{amsfonts}
\usepackage{amssymb}
\usepackage{graphicx}
\usepackage{textcomp}
\usepackage{color}
\begin{document}

\title{Magnetic properties of the finite-length biatomic chains in the framework of the single domain-wall approximation}

\author{S.V. Kolesnikov}
\email{kolesnikov@physics.msu.ru}
\affiliation{Faculty of Physics, Lomonosov Moscow State University, Moscow 119991, Russian Federation}
\author{I.N. Kolesnikova}
\affiliation{Department of Chemistry, Lomonosov Moscow State University, Moscow 119991, Russian Federation}

\begin{abstract}
A simple analytical method for study the magnetic properties of the finite-length biatomic chains in the framework of Heisenberg model with uniaxial magnetic anisotropy is proposed. The method allows to estimate the reversal time of the magnetization of ferromagnetic and antiferromagnetic biatomic chains. Three cases are considered: the spontaneous remagnetization, the remagnetization under the interaction with a scanning tunneling microscope, and the remagnetization in the external magnetic field. The applicability limits of the method are discussed. Within its limits of applicability the method gives results which are in perfect agreement with the results of the kinetic Monte Carlo simulations. As the examples, two physical systems are considered: biatomic Fe chains on Cu$_2$N/Cu(001) surface and biatomic Co chains on Pt(997) surface. The presented method is incomparably less time-consuming than the standard kMC simulations, especially in the cases of low temperatures or long chains.
\end{abstract}	


\keywords{biatomic chains, magnetic properties, Heisenberg model, single domain-wall approximation}

\date{\today}
	
\maketitle

\section{Introduction}\label{INTRODUCTION}

The investigations of the magnetic properties of the atomic chains are of general interest due to the prospects for the creation of the next generation mass storage devices~\cite{JPCM16.R603,JPCM22.433001,NANO6.1}. For application of the atomic chains as bits of information, its reversal time of the magnetization needs to be sufficiently long. The possibility of engineering of such memory elements~\cite{Nature437.671} appeared after the discovery of the giant magnetic anisotropy energy (MAE) of Co atoms on the  Pt(997) surface~\cite{Gambardella.Nature,Gambardella_2003} using X-ray magnetic circular dichroism (XMCD) and scanning tunneling microscope (STM)~\cite{Brune.Gambardella,PRL102.257203,SSR56.189}. Ferromagnetic Co chains grow on the step edges of Pt(997) surface at low concentrations of Co atoms. The analogous effect was observed for Fe/Cu(111) system~\cite{JPCM15.R1,PRB56.2340}. The critical temperature $T_\text{C}$ and the reversal time of the magnetization $\tau$ of the atomic chain increase with their length. According to the estimation~\cite{Gambardella_2003} the atomic chain consisting of 400 Co atoms can be a stable bit at room temperature. To increase the information recording density, it is possible to use biatomic ferromagnetic chains~\cite{PRL93.077203,NJPhys17.023014}. However, an increase in the width of the chain can lead to the significant decrease of MAE~\cite{PRL93.077203,PRB89.205426,NJP4.100}. These observations are in a good agreement with the well-known effect of decreasing of the average MAE of atoms in atomic clusters with an increase of its size~\cite{JPCM28.503002,PRB70.224419,JETPL99.646}.

Another interesting opportunity is the use of finite-size antiferromagnetic chains as bits of information~\cite{NuturePhys14.213,RevModPhys90.015005,LTP40.17,NatureNanotech11.231}. The interaction between antiferromagnetic chains is much weaker than the interaction between ferromagnetic ones. Therefore, the use of antiferromagnetic chains can lead to a significant increase in the information recording density. The possibility of creating and remagnetization of such chains using STM was demonstrated by the example of Fe atomic chains  on Cu$_2$N/Cu(001) surface~\cite{science335.196,NatureNano10.40}. A systematic study of the transition metal atomic chains on Cu$_2$N/Cu(001) surface has shown that its can be both ferromagnetic and antiferromagnetic~\cite{PRB92.184406,PRB92.174407,PRB94.085406,PRB86.245416,PCCP17.26302}. Very similar results are obtained for the atomic chains on Cu$_2$O/Cu(001) surface~\cite{PRB90.195423}. Biatomic antiferromagnetic chains are significantly more  stable than the single-atomic ones~\cite{science335.196}. Special attention should be paid to the investigations of the remagnetization of the atomic chains with the STM tip. It was shown that high STM voltage transitions mediated by domain-wall formation~\cite{PRL110.087201}.

A lot of theoretical investigations are devoted to the ferromagnetic and antiferromagnetic finite-size chains. Among them it is necessary to underline the studies of influence of quantum tunneling on the reversal time of the magnetization~\cite{JPCM27.455301,EPL109.57001,PRL110.087201}. Quantum tunneling is recovered as the switching mechanism at extremely low temperatures below the mK range for a six-Fe-atom system and exponentially lower for larger atomic systems~\cite{JPCM27.455301}. Therefore, one can neglect the quantum nature of the magnetic moments of atoms in a wide range of temperatures. In this case the magnetic properties of atomic chains can be described in the framework of the classical Heisenberg Hamiltonian and its generalizations.

The parameters of the Heisenberg Hamiltonian can be calculated from the first principles by means of density functional theory~\cite{PRB90.195423,NJPhys17.023014,PRB93.161412R} or Korringa–Kohn–Rostoker-Green’s function method~\cite{RPP74.096501,PRL102.257203}. Further investigation of the magnetic properties of the chains can be performed with either the solution of the Landau–Lifshitz-Gilbert equation~\cite{Landau,PRB93.161412R} or the kinetic Monte Carlo (kMC) simulations~\cite{PhysRevB.73.174418}. The kMC method allows to calculate the critical temperature, the reversal time of the magnetization, and the coercive field of the ferromagnetic chains~\cite{JMMM378.186,NJPhys11.063004,PRL96.217201,CPB24.097302,PSS57.1513}. Dynamical magnetization of rectangular lattices of interacting tiny magnets with strong perpendicular uniaxial anisotropy was also investigated~\cite{PhysLettA374.2058}. The kMC method can be successfully applied to the study of the antiferromagnetic chains~\cite{PRB93.035444,MPLB.31.1750142}.

However, the kMC method is a statistical method. Obtaining of the averaged values with small errors always needs a lot of simulations.
Usual kMC simulations can be very  time-consuming, especially in the cases of low temperatures or long chains. Thus, it would be useful to have a simlpe analytical method for estimation of the reversal time of the magnetization, which would be in a good agreement with the results of the kMC simulations in a wide range of parameters. Such method in the single domain-wall approximation was developed for the single-atomic chains~\cite{Kolesnikov1,Kolesnikov2}. It was shown that the single domain-wall approximation is justified in a wide range of parameters of the Heisenberg Hamiltonian and a wide range of temperatures. The proposed method allows to estimate the reversal time of the magnetization of ferromagnetic chains in the cases of the spontaneous remagnetization and the remagnetization in the external magnetic field~\cite{Kolesnikov1}. The reversal time of the magnetization of antiferromagnetic chains in the case of the remagnetization under the interaction with the STM tip also can be estimated~\cite{Kolesnikov2}.

In this article, we generalize the previously developed formalism~\cite{Kolesnikov1,Kolesnikov2} to the case of the biatomic chains. We will consider two limiting cases: a weak and a strong coupling between the atomic chains. We will see that these two approximations cover a wide range of parameters and can be used in a lot of practically interesting situations. The estimation formulas for the reversal time of the magnetization of the biatomic chains will be derived in three following cases: (i) the spontaneous remagnetization, (ii) the remagnetization under the interaction with STM tip, and (iii) the remagnetization in the external magnetic field. We will show that our method allows to calculate the magnetization curves and the coercive fields of biatomic ferromagnetic chains. The obtained analytical results will be compared with the results of kMC simulation.

The paper is organized as follows. In Section~\ref{THEORY}, the theoretical model is briefly discussed. In Section~\ref{free_chain}, we derive the estimation formulas for the reversal time of the magnetization in the case of the spontaneous remagnetization of both ferromagnetic and antiferromagnetic biatomic chains. Interaction of biatomic chains with STM is discussed in Section~\ref{STM_interaction}.
Remagnetization of ferromagnetic biatomic chains in the external magnetic field is investigated in Section~\ref{B_field}. In order to demonstrate the ability of our method to estimate the reversal time of the magnetization some numerical results are compared with the results of the kMC simulations in Section~\ref{Numerical}. We conclude the paper in Section~\ref{conc}. For the reader's convenience, the main results from~\cite{Kolesnikov1, Kolesnikov2} are summarized in Appendix.

\section{Theoretical model}\label{THEORY}

In order to estimate the reversal time of the magnetization we assume the following. We can neglect quantum tunneling at the temperatures $T>T_\text{QT}$ and consider the magnetic moments of atoms as classical vectors. Temperature $T_\text{QT}$ has an order of mK for the chains under consideration~\cite{JPCM27.455301}. Following the work of Li and Liu~\cite{PhysRevB.73.174418} we consider the case of uniaxial magnetic anisotropy. Thus the Heisenberg Hamiltonian can be written in the following form
\begin{equation}\label{eq2}
H=-\sum_{i>j}J_{ij}\left({\bf s}_i\cdot{\bf s}_j\right)-K\sum_{i}\left({\bf s}_i\cdot{\bf e}\right)^2
-\mu\sum_{i}\left({\bf s}_i\cdot{\bf B}\right),
\end{equation}
where ${\bf s}_i$ and ${\bf e}$ are the unit vectors of the magnetic moments of the atoms and the easy axis of magnetization, respectively, $\mu$ is the absolute value of the magnetic moments, $K$ is MAE, $J_{ij}=J(\delta_{i,j+1}+\delta_{i,j-1})$ is the exchange energy, $\delta_{ij}$ is Kronecker delta. For the ferromagnetic chains $J>0$ and for the antiferromagnetic chains $J<0$. The external magnetic field ${\bf B}$ is assumed to be applied along the easy axis of magnetization ${\bf e}$. We assume that all of the magnetic moments are directed either parallel or antiparallel to the easy axis of magnetization $\left({\bf s}_i\cdot {\bf e}\right)=\pm1$. We can say that the magnetic moment is directed ``up'' if $\left({\bf s}_i\cdot{\bf e}\right)=1$, and ``down'' if $\left({\bf s}_i\cdot{\bf e}\right)=-1$.

The flipping of the $i$-th magnetic moment can occur in two different ways. If $2K>|h_i|$ (where $h_i=\sum_{j}J_{ij}({\bf s}_i\cdot{\bf s}_j)+\mu({\bf s}_i\cdot{\bf B})$) then the rate of the single magnetic moment flip is determined as
\begin{equation}\label{eq4}
\nu(h_i)=\nu_0\exp\left(-\frac{\left(2K+h_i\right)^2}{4Kk_\text{B}T}\right),
\end{equation}
where $k_\text{B}$ is the Boltzmann constant, $T$ is the temperature, and $\nu_0$ is the frequency prefactor. If $2K\le|h_i|$, then there is no energy barrier between the states $\left({\bf s}_i\cdot{\bf e}\right)=\pm1$. The rate of the single magnetic moment flip can be calculated~\cite{Glauber1.1703954}, as
\begin{equation}\label{eq5}
\nu(h_i)=\nu_0\frac{\exp(-2h_i/k_\text{B}T)}{1+\exp(-2h_i/k_\text{B}T)}.
\end{equation}
Below we will use the function $\nu(h_i)$, which is given by formulas (\ref{eq4}) and (\ref {eq5})~\footnote{This function is discontinuous. Instead of (\ref{eq5}) one can use, for example, the expression $\nu(h_i)=\nu_0\min\left(1,\exp(-2h_i/k_\text{B}T)\right)$. Then the function $\nu(h_i)$ will be continuous. In any case, it does not affect the applicability limits of our model, because we use the same functions $\nu(h_i)$ for analytical estimates and the kMC simulations.}. The frequency prefactor $\nu_0=10^9$~Hz~\cite{Gambardella.Nature} is chosen for the numerical estimates.

\begin{figure}[htb]
\begin{center}
\includegraphics[width=0.7\linewidth]{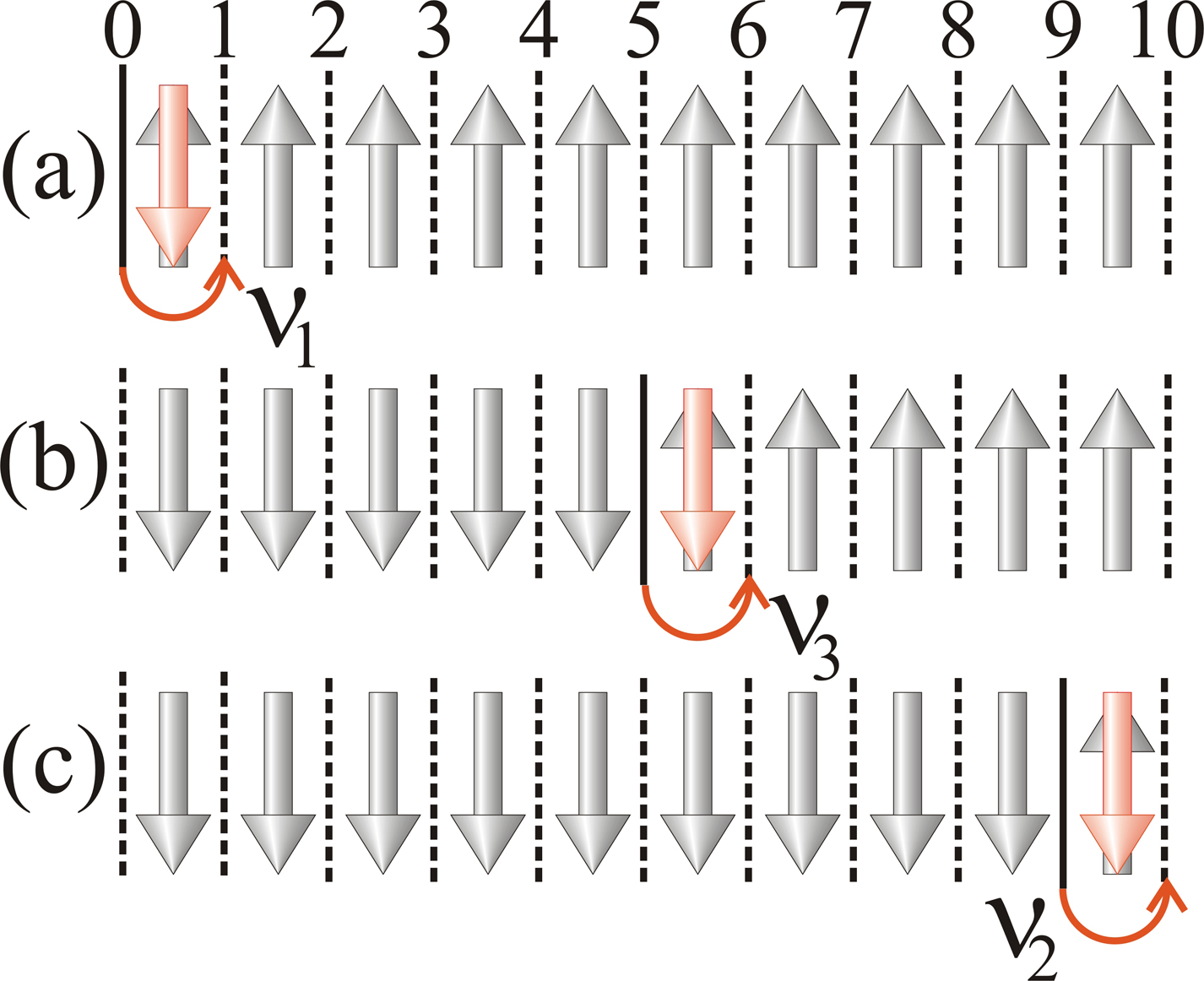}
\caption{\label{fig1} A schematic view of the atomic chain consisting of $N=10$ magnetic moments. Dashed lines show the possible positions $i=0,\dots,10$ of the domain wall. Solid lines show its current position. The following processes are shown: (a) formation of the domain wall with the rate of $\nu_1$, (b) motion of the domain wall with the rate of $\nu_3$, and (c) the domain wall disappearance with the rate of $\nu_2$. The red arrows show the rotating magnetic moments.
}
\end{center}
\end{figure}

Here we discuss the main ideas of our method on the example of the ferromagnetic chain consisting of 10 atoms. Let all of the magnetic moments be directed up in the initial moment of time (Fig.~\ref{fig1}). We assume that the remagnetization of the chain is associated with the formation and motion of the single domain wall. The width of the domain wall can be neglected. Dashed lines in Fig.~\ref{fig1} show the possible positions $i=0,\dots,10$ of the domain wall. Solid lines show its current position: (a) $i=0$, (b) $i=5$, (c) $i=9$. The domain wall leaves initial state $i=0$, moves randomly along the chain, and comes to final state $i=10$. We define the reversal time of the magnetization $\tau$ of the atomic chain as the average time of the random walk of the domain wall. If the parameters of the Hamiltonian~(\ref{eq2}) are the same for all of the atoms and ${\bf B}=0$ then the random walk of the domain wall is characterized by only three rates: (i) the rate of formation of the domain wall $\nu_1$ (Fig.~\ref{fig1}(a)), (ii) the rate of the domain wall disappearance $\nu_2$ (Fig.~\ref{fig1}(c)), and (iii) the rate of motion of the domain wall along the chain $\nu_3$ (Fig.~\ref{fig1}(b)). The rate $\nu_{1,2,3}$ can be easily calculated using the formulas~(\ref{eq4}) and~(\ref{eq5}). We assume that two domain walls can not exist simultaneously if the temperature is lower than $T_\text{max}$. It is obvious that $T_\text{max}<T_\text{C}$. Thus, our model is valid in the temperature range of $T_\text{QT}<T<T_\text{max}$.

To calculate the average time of a random walk of the domain wall the mean rate method is employed~\cite{PMA76.565,JCP132.134104}. At first, we need to calculate the rates $\nu_{i\to j}$ of all of the possible transitions of the domain wall and find the transition probability matrix
\begin{equation}\label{eq15}
T_{ij}=\tau^1_j\nu_{j\to i},
\end{equation}
where $\tau^1_j=\left(\sum_k \nu_{j\to k}\right)^{-1}$ is the mean residence time in state $j$ each time it is occupied, indexes $i$ and $j$ run over initial and all of transient states of the domain wall, index $k$ runs over all possible states (including final states) of the domain wall. The probability $P_i$ of finding the domain wall in state $i$ can be found from the following system of linear equations
\begin{equation}\label{eq18}
\sum\limits_{j=0}^{N-1}\left(\delta_{ij}-T_{ij}\right)P_j=P^{\text{init}}_i,
\end{equation}
where $P^{\text{init}}_i=\delta_{0i}$ is the probability of finding the domain wall in state $i$ at the initial moment of time. The average time of a random walk of the domain wall can be obtained as
\begin{equation}\label{eq24}
\tau_{\text{tot}}=\sum\limits_{i=0}^{N-1}\tau^1_iP_i.
\end{equation}
If the remagnetization of the chain can begin from any of its ends with the same probability, then its reversal time of the magnetization $\tau$ is equal to $\tau_{\text{tot}}/2$.

\section{Results and Discussions}\label{RES_AND_DISC}

\begin{figure}[htb]
\begin{center}
\includegraphics[width=0.95\linewidth]{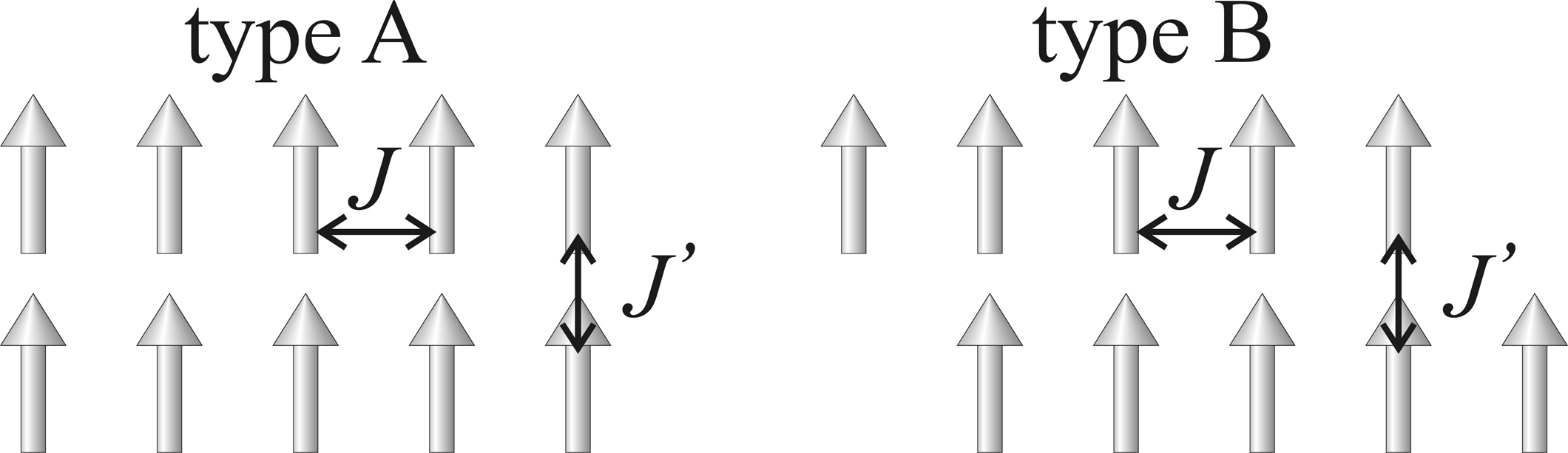}
\caption{\label{fig2} Two types of the biatomic chains under consideration: (a) type A and (b) type B. Exchange energies $J$ and $J'$ characterize the coupling between the neighboring atoms.
}
\end{center}
\end{figure}

Following the experimental work~\cite{science335.196} we consider biatomic chains of two types: type A and type B (see Fig.~\ref{fig2}). We assume that the exchange energy $J$ characterizes the interactions between the nearest atoms in the same atomic chain, and $J'$ characterizes the interactions between the atoms of neighboring atomic chains. For simplicity, the interactions between other pairs of atoms are neglected. The all of rates $\nu(h_i)$ which will be used below can be calculate by means of formulas (\ref{eq4}) and (\ref{eq5}).
We will need the following rates:
$\nu_{1}=\nu(|J|+|J'|)$,
$\nu'_{1}=\nu(|J|-|J'|)$,
$\nu''_{1}=\nu(|J|)$,
$\nu_{2}=\nu(-|J|+|J'|)$,
$\nu'_{2}=\nu(-|J|-|J'|)$,
$\nu''_{2}=\nu(-|J|)$,
$\nu_{3}=\nu(|J'|)$,
$\nu'_{3}=\nu(-|J'|)$,
$\nu''_{3}=\nu(0)$ if ${\bf B}=0$,
and
$\nu_{1\pm}=\nu(J+J'\pm\mu B)$,
$\nu'_{1\pm}=\nu(J-J'\pm\mu B)$,
$\nu''_{1\pm}=\nu(J\pm\mu B)$,
$\nu_{2\pm}=\nu(-J+J'\pm\mu B)$,
$\nu'_{2\pm}=\nu(-J-J'\pm\mu B)$,
$\nu''_{2\pm}=\nu(-J\pm\mu B)$,
$\nu_{3\pm}=\nu(J'\pm\mu B)$,
$\nu'_{3\pm}=\nu(-J'\pm\mu B)$,
$\nu''_{3\pm}=\nu(\pm\mu B)$ if ${\bf B}\ne0$.
Below, we derive formulas for estimating the reversal time of the magnetization of the biatomic chains in two limiting cases: (i) a weak coupling between the atomic chains ($|J'|\ll |J|$) and (ii) a strong coupling between the chains ($|J'|\gtrsim|J|$). The parameters of the Heisenberg Hamiltonian are assumed to be the same for all of the atoms. Edge effects can be taken into account as well as in the case of the single-atomic chains (see~\cite{Kolesnikov2} and formulas (\ref{eq21a}) and (\ref{eq22a}) in Appendix).

\subsection{The spontaneous remagnetization}\label{free_chain}

We begin our investigation with the case of the spontaneous remagnetization of the biatomic chains in the absence of external fields (no interaction with the STM tip, no external magnetic field ${\bf B}=0$). How it can be seen below, all of the formulas for the reversal times of the magnetization include the absolute values of the exchange energies $|J|$ and $|J'|$. Thus, all of the results obtained in this Section are valid for both ferromagnetic and antiferromagnetic chains.

\subsubsection{A weak coupling approximation}

\begin{figure}[htb]
\begin{center}
\includegraphics[width=0.7\linewidth]{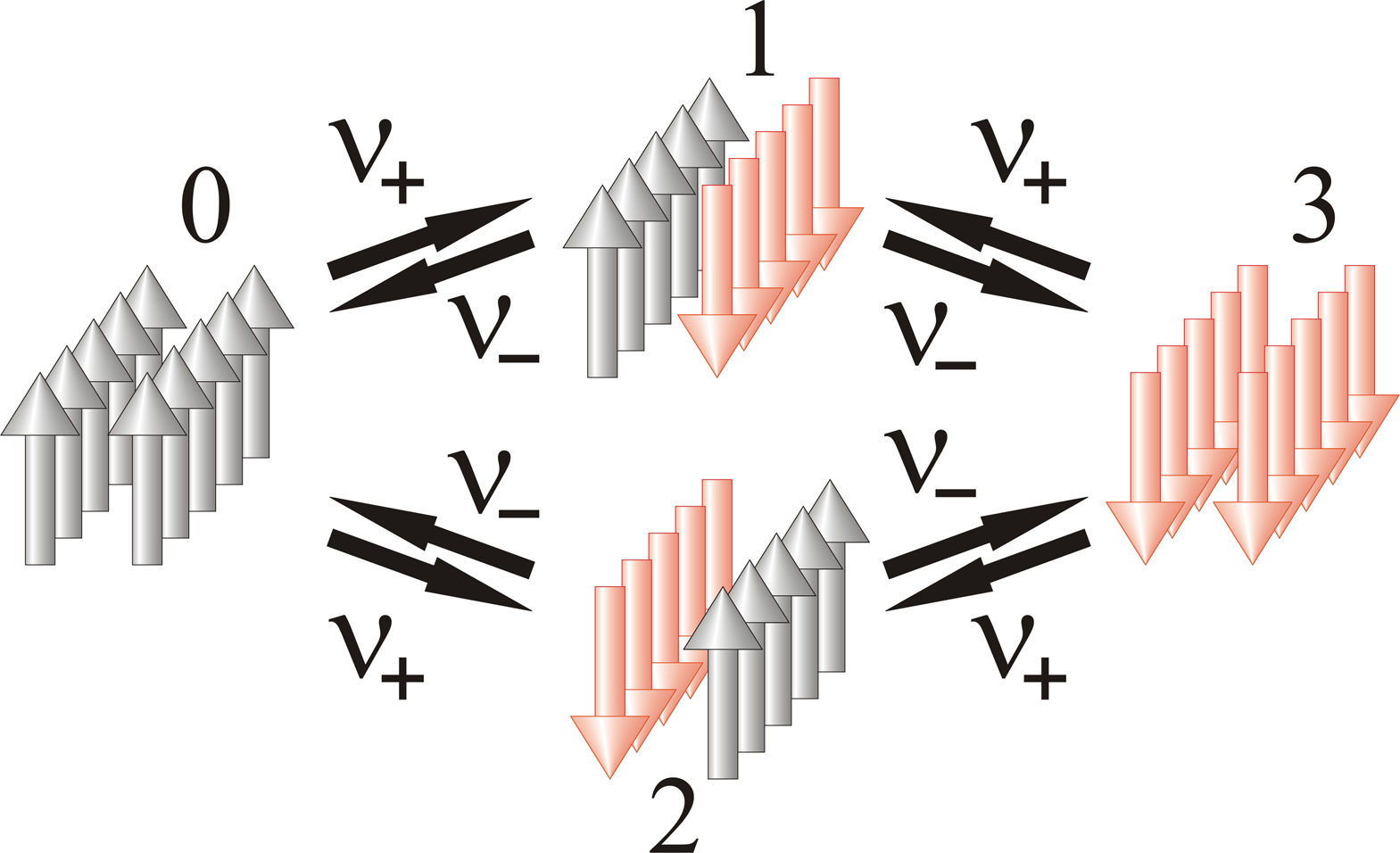}
\caption{\label{fig3} A schematic view of the remagnetization of the biatomic chain in a weak coupling approximation: 0 is the initial state, 1 and 2 are the transient states, and 3 is the final state. The red arrows show the rotating magnetic moments.
}
\end{center}
\end{figure}

Let us consider weakly interacting atomic chains ($|J'|\ll |J|$). This case is directly related to the experimental work~\cite{science335.196}, because the ratio between the exchange energies of the Fe atoms on Cu$_2$N/Cu(001) surface is $J/J'\approx40$. We assume that the atomic chains flip one by one (see Fig.~\ref{fig3}). So, the atomic chain flipping at the current moment is in the effective magnetic field which is created by another chain. The initial state, two transient states, and the final state of the biatomic chain are denoted as 0, 1, 2, and 3, respectively. Then
\begin{equation}\label{eq1b}
\nu_{0\to 1}=\nu_{0\to 2}=\nu_{3\to 1}=\nu_{3\to 2}=\nu_{+},
\end{equation}
\begin{equation}\label{eq2b}
\nu_{1\to 0}=\nu_{2\to 0}=\nu_{1\to 3}=\nu_{2\to 3}=\nu_{-},
\end{equation}
\begin{equation}\label{eq3b}
\nu_{0\to 3}=\nu_{3\to 0}=0.
\end{equation}
The non-zero elements of the transition probability matrix calculated by the formula (\ref{eq15}) are equal to $T_{0i}=1/2$, $T_{i0}=1/2$, where $i=1,2$. After solving the system of equations (\ref{eq18}) with $P^{\text{init}}=\{1,0,0\}$ we find the reversal time of the magnetization in the weak coupling approximation
\begin{equation}\label{eq5b}
\tau^\text{weak}=\tau_{+}+\tau_{-},
\end{equation}
where $\tau_{+}=1/\nu_{+}$ and $\tau_{-}=1/\nu_{-}$.

The reversal times of the magnetization $\tau_{+}$ and $\tau_{-}$ are different for the biatomic chains of type A and type B. For the chains of type A, all of the atoms of the second atomic chain are in the effective magnetic field $B=|J'|/\mu$ created by the atoms of the first atomic chain at the transition $0\to1$ (see Fig.\ref{fig3}). Therefore, the reversal time of the magnetization of the second atomic chains can be estimated by the formula (\ref{eq38}) in which we make the following replacements:
$\nu''_{1+}\to\nu_1$, $\nu''_{1-}\to\nu'_1$,
$\nu''_{2+}\to\nu_2$, $\nu''_{2-}\to\nu'_2$,
$\nu''_{3+}\to\nu_3$, $\nu''_{3-}\to\nu'_3$, i.e.
\begin{equation}\label{eq6b}
\tau_{+}=\tau_\text{B}\left(\frac{|J'|}{\mu}\right).
\end{equation}
All of the atoms of the first atomic chain are in the effective magnetic field $B=-|J'|/\mu$ created by the atoms of the second atomic chain at the transition $1\to3$.  Thus,
\begin{equation}\label{eq7b}
\tau_{-}=\tau_\text{B}\left(-\frac{|J'|}{\mu}\right).
\end{equation}
Replacing $|J'|\to-|J'|$ is equivalent to replacing $\nu_{i}\to\nu'_{i}$, $\nu'_{i}\to\nu_{i}$ in formula (\ref{eq6b}), where $i=1,2,3$.

In the case of the biatomic chain of type B, one of the edge atoms of each atomic chains does not interact with atoms of another chain. In other words, one of the ends of each atomic chain is ``free''. Let us consider the transition $0\to1$. Now the reversal time of the magnetization of the second atomic chain depends on which end the remagnetization starts from. If the remagnetization begins at the free end, then the reversal time of the magnetization can be estimated by the formula (\ref{eq38}) after the replacements $\nu''_{1+}\to\nu''_1$, $\nu''_{2-}\to\nu''_2$ and multiplying by factor of 2:
\begin{equation}\label{eq8b}
\tau_{1+}=2\tau_\text{B}\left(\frac{|J'|}{\mu};\nu''_{1+}\to\nu''_1,\nu''_{2-}\to\nu''_2\right).
\end{equation}
If the remagnetization of the second atomic chain begins at another end, then the reversal time of the magnetization is the following
\begin{equation}\label{eq9b}
\tau_{2+}=2\tau_\text{B}\left(\frac{|J'|}{\mu};\nu''_{1-}\to\nu''_1,\nu''_{2+}\to\nu''_2\right).
\end{equation}
Then the average reversal time of the magnetization of the second chain is equal to
\begin{equation}\label{eq10b}
\tau_{+}=\left(\frac{1}{\tau_{1+}}+\frac{1}{\tau_{2+}}\right)^{-1}.
\end{equation}
We find the similar result for the transition $1\to3$:
\begin{equation}\label{eq11b}
\tau_{-}=\left(\frac{1}{\tau_{1-}}+\frac{1}{\tau_{2-}}\right)^{-1},
\end{equation}
where
\begin{equation}\label{eq12b}
\tau_{1-}=2\tau_\text{B}\left(-\frac{|J'|}{\mu};\nu''_{1+}\to\nu''_1,\nu''_{2-}\to\nu''_2\right),
\end{equation}
\begin{equation}\label{eq13b}
\tau_{2-}=2\tau_\text{B}\left(-\frac{|J'|}{\mu};\nu''_{1-}\to\nu''_1,\nu''_{2+}\to\nu''_2\right).
\end{equation}

Finally, we find the reversal time of the magnetization $\tau^\text{weak}$ of the biatomic chains in a weak coupling approximation by substituting either (\ref{eq6b}) and (\ref{eq7b}) for the chains of type A, or (\ref{eq10b}) and (\ref{eq11b}) for the chains of type B. Note that in the limit of $|J'|\to0$ the value of $\tau^\text{weak}$ tends to $2\tau_1$, where $\tau_1$ is the reversal time of the magnetization of the single-atomic chain in the absence of external fields calculated by the formula (\ref{eq25}).

\subsubsection{A strong coupling approximation}

\begin{figure}[htb]
\begin{center}
\includegraphics[width=0.7\linewidth]{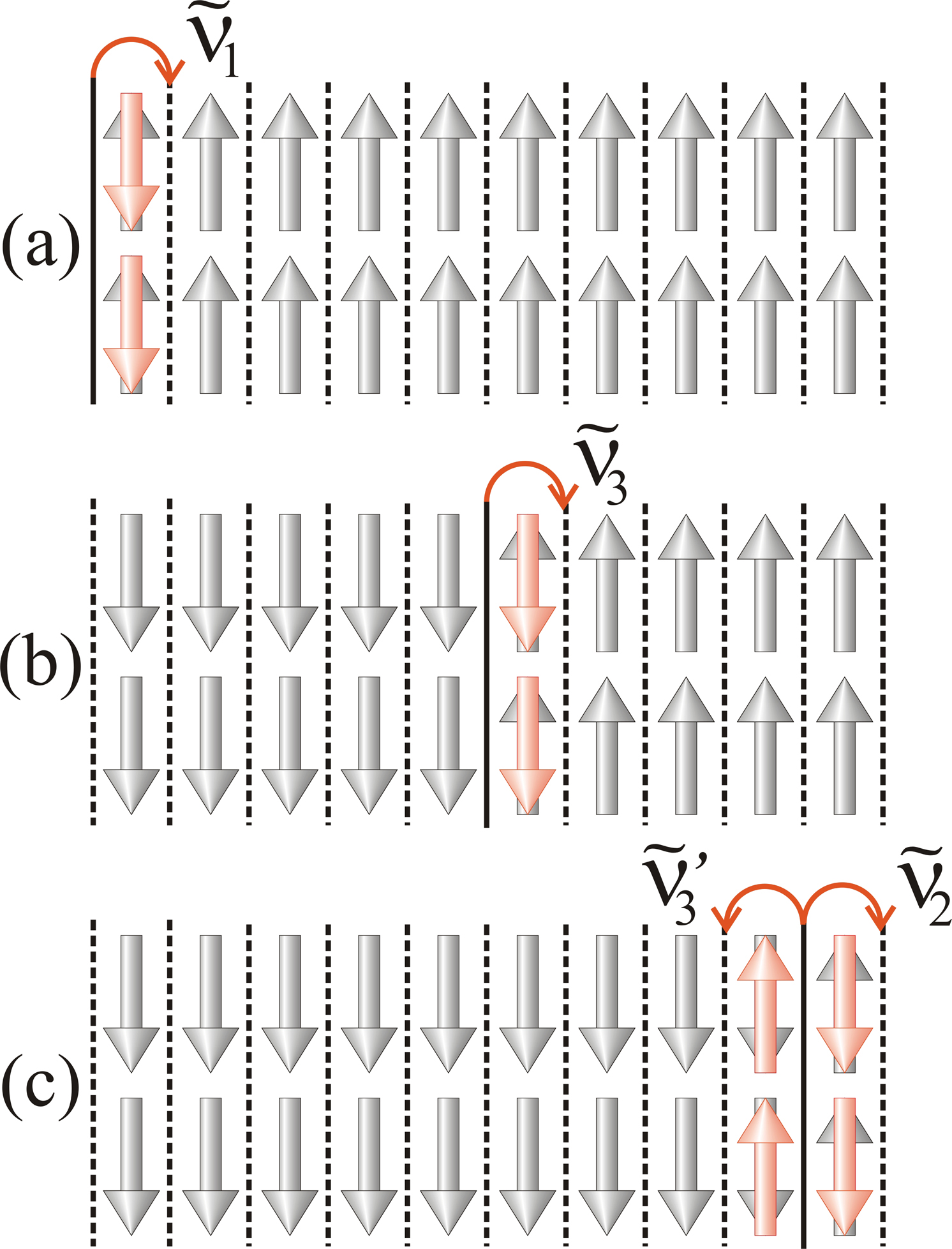}
\caption{\label{fig4} A schematic view of the biatomic chain of type A consisting of $2N=20$ magnetic moments. Dashed lines show the stable positions of the domain wall in the strong coupling approximation. Solid lines show its current position. The following processes are shown: (a) formation of the domain wall with the rate of $\tilde\nu_1$, (b) motion of the domain wall with the rate of $\tilde\nu_3$, and (c) the domain wall disappearance with the rate of $\tilde\nu_2$ and motion of the domain wall near the end of the chain with the rate of $\tilde\nu'_3$. The red arrows show the rotating magnetic moments.
}
\end{center}
\end{figure}

In the case of a strong coupling between the atomic chains ($|J'|\gtrsim|J|$) the length of the domain wall should be minimal. So, the domain wall must be perpendicular to the biatomic chain. We first consider the simpler case of the biatomic chain of type A.
The positions of the domain wall corresponding to the local minima of the energy are shown in Fig.~\ref{fig4}. To estimate the reversal time of the magnetization of the biatomic chain, it is necessary to calculate the rates $\tilde\nu_1$, $\tilde\nu_2$, $\tilde\nu_3$, and $\tilde\nu'_3$ of formation, disappearance and motion of the domain wall.

\begin{figure}[htb]
\begin{center}
\includegraphics[width=0.9\linewidth]{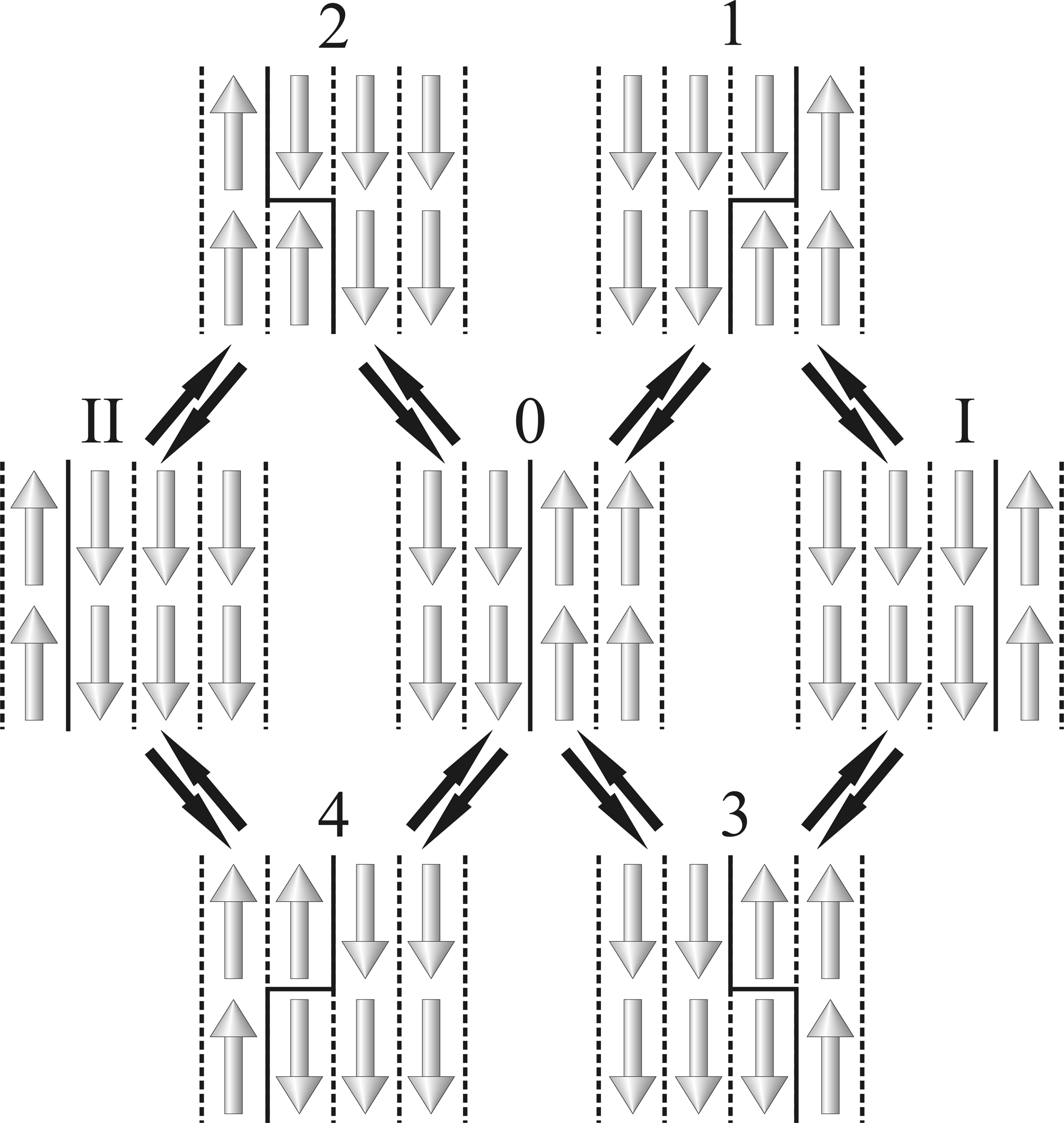}
\caption{\label{fig5} A schematic view of motion of the domain wall in a strong coupling approximation: 0 is the initial state, 1, 2, 3, and 4 are the transient states, I and II are the final states. The red arrows show the rotating magnetic moments.
}
\end{center}
\end{figure}

Let us calculate the rate of motion of the domain wall $\tilde\nu_3$. Figure~\ref{fig5} shows that the domain wall can transit from the initial state 0 to one of two equivalent final states I or II through the transient states 1,2,3,4. The non-zero transition rates are equal to
\begin{equation}\label{eq14b}
\nu_{0\to i}=\nu_3,
\end{equation}
\begin{equation}\label{eq15b}
\nu_{i\to 0}=\nu_{1\to I}=\nu_{2\to II}=\nu_{3\to I}=\nu_{4\to II}=\nu'_3,
\end{equation}
where $i=1,2,3,4$. The non-zero elements of the transition probability matrix calculated by the formula (\ref{eq15}) are equal to $T_{0i}=1/2$, $T_{i0}=1/4$. After solving the system of equations (\ref{eq18}) with $P^{\text{init}}=\{1,0,0,0,0\}$ we find
\begin{equation}\label{eq16b}
\tilde\nu_{3}=\frac{\tau_1^1P_1\nu_{1\to I}+\tau_3^1P_3\nu_{3\to I}}{\tau_\text{tot}}=\frac{\nu_3\nu'_3}{\nu'_3+2\nu_3}.
\end{equation}

\begin{figure}[htb]
\begin{center}
\includegraphics[width=0.7\linewidth]{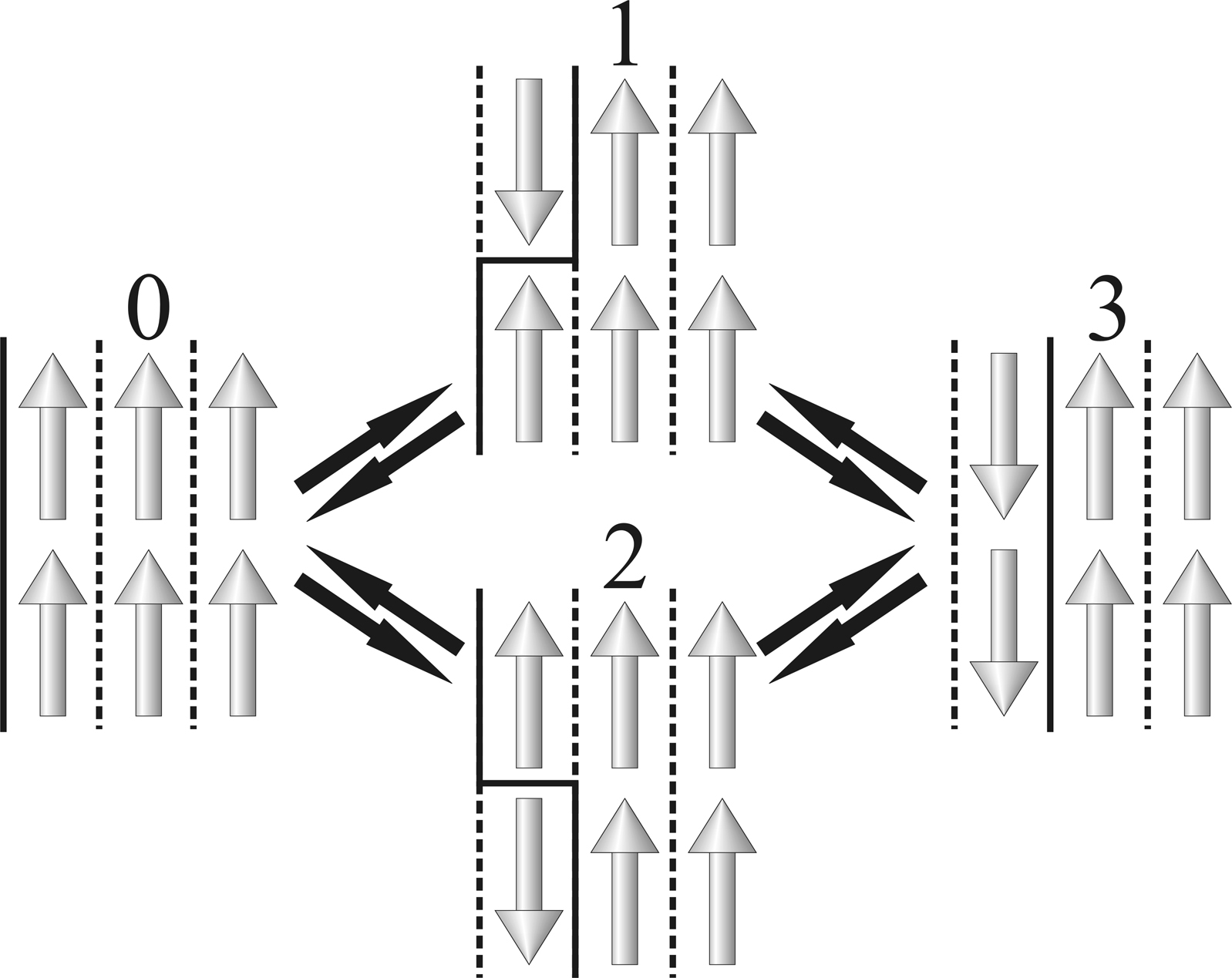}
\caption{\label{fig6} A schematic view of formation of the domain wall in the chain of type A in a strong coupling approximation: 0 is the initial state, 1 and 2 are the transient states, 3 is the final state. The red arrows show the rotating magnetic moments.
}
\end{center}
\end{figure}

Now let us calculate the formation rate of the domain wall $\tilde\nu_1$. Figure~\ref{fig6} shows that the domain wall can transit from the initial state 0 to the final state 3 through the transient states 1 or 2. The non-zero transition rates are equal to
\begin{equation}\label{eq17b}
\nu_{0\to i}=\nu_1,~~~~~\nu_{i\to3}=\nu'_1,~~~~~\nu_{i\to 0}=\nu'_2,
\end{equation}
where $i=1,2$. The non-zero elements of the transition probability matrix are equal to $T_{0i}=\nu'_2/(\nu'_1+\nu'_2)$, $T_{i0}=1/2$. After solving the system of equations (\ref{eq18}) with $P^{\text{init}}=\{1,0,0\}$ we find
\begin{equation}\label{eq18b}
\tilde\nu_{1}=\frac{\tau_1^1P_1\nu_{1\to 3}+\tau_2^1P_2\nu_{2\to 3}}{\tau_\text{tot}}=\frac{2\nu_1\nu'_1}{\nu_1+\nu'_1+\nu'_2}.
\end{equation}

\begin{figure}[htb]
\begin{center}
\includegraphics[width=0.85\linewidth]{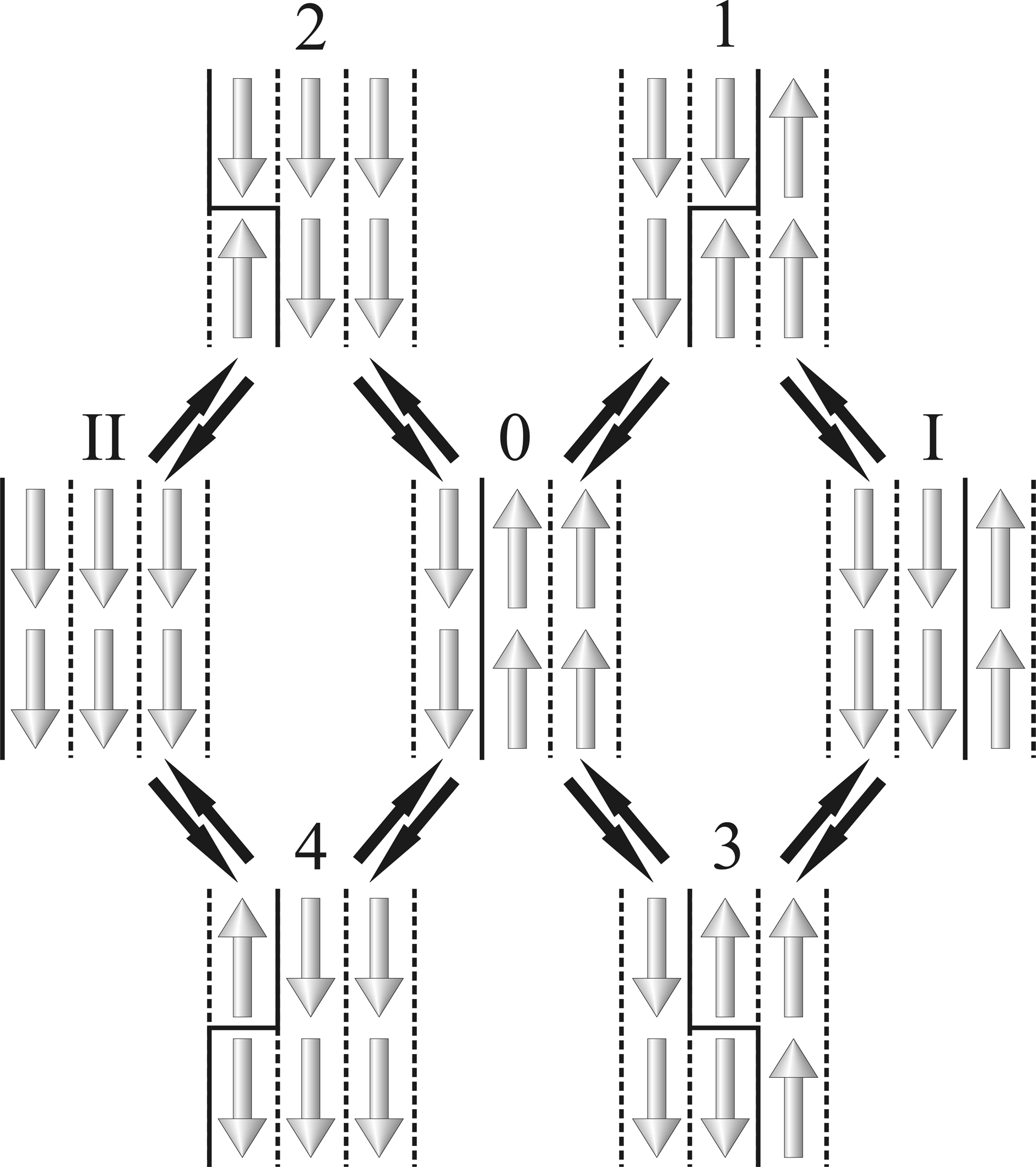}
\caption{\label{fig7} A schematic view of the domain wall disappearance and motion of the domain wall near the end of the chain of type A in a strong coupling approximation: 0 is the initial state, 1, 2, 3, and 4 are the transient states, I and II are the final states. The red arrows show the rotating magnetic moments.
}
\end{center}
\end{figure}

Finally, let us calculate the rate of the domain wall disappearance $\tilde\nu_2$ and the rate $\tilde\nu'_3$. The domain wall can transit from the initial state 0 to one of two nonequivalent final states I or II through the transient states 1,2,3,4 (see Fig.~\ref{fig7}). The non-zero transition rates are equal to
\begin{equation}\label{eq19b}
\nu_{0\to 1}=\nu_{0\to 3}=\nu_3,~~~~~\nu_{0\to 2}=\nu_{0\to 4}=\nu_2,
\end{equation}
\begin{equation}\label{eq20b}
\nu_{1\to 0}=\nu_{3\to 0}=\nu_{1\to I}=\nu_{3\to I}=\nu'_3,
\end{equation}
\begin{equation}\label{eq21b}
\nu_{2\to 0}=\nu_{4\to 0}=\nu'_1,~~~~~\nu_{2\to II}=\nu_{4\to II}=\nu'_2.
\end{equation}
The non-zero elements of the transition probability matrix are equal to $T_{01}=T_{03}=1/2$, $T_{02}=T_{04}=\nu'_1/(\nu'_1+\nu'_2)$, $T_{10}=T_{30}=a/2$, $T_{20}=T_{40}=(1-a)/2$, where $a=\nu_3/(\nu_2+\nu_3)$. After solving the system of equations (\ref{eq18}) with $P^{\text{init}}=\{1,0,0,0,0\}$ we find
\begin{multline}\label{eq22b}
\tilde\nu_{2}=\frac{\tau_2^1P_2\nu_{2\to II}+\tau_4^1P_4\nu_{4\to II}}{\tau_\text{tot}}=\\
=\frac{2\nu_2\nu'_2\nu'_3}{\nu_3(\nu'_1+\nu'_2)+\nu'_3(2\nu_2+\nu'_1+\nu'_2)},
\end{multline}
\begin{multline}\label{eq23b}
\tilde\nu'_{3}=\frac{\tau_1^1P_1\nu_{1\to I}+\tau_3^1P_3\nu_{3\to I}}{\tau_\text{tot}}=\\
=\frac{\nu_3\nu'_3(\nu'_1+\nu'_2)}{\nu_3(\nu'_1+\nu'_2)+\nu'_3(2\nu_2+\nu'_1+\nu'_2)}.
\end{multline}

Now the problem of the remagnetization of the biatomic chain is reduced to the problem of the remagnetization of a single-atomic chain. In order to estimate the reversal time of the magnetization $\tau^\text{strong}$, we can use the formula (\ref{eq22a}) where we should replace $\nu''_1\to\tilde\nu_1$, $\nu''_2\to\tilde\nu_2$, $\nu_2,\nu''_3\to\tilde\nu_3$, $\nu'_1\to\tilde\nu'_3$. Then
\begin{multline}\label{eq24b}
\tau^\text{strong}=\frac{1}{2\tilde a}\left\{\frac{\tilde a}{2\tilde\nu_3}
\left(N-3+2\frac{\tilde\nu_3}{\tilde\nu'_3}\right)
\left[N-\frac{2(1-2\tilde a)}{1-\tilde a}\right]\right.+\\
+\left.\frac{1}{\tilde\nu_1}\left[N(1-\tilde a)-2(1-2\tilde a)\right]\right\},
\end{multline}
where $\tilde a=\tilde\nu'_3/(\tilde\nu_2+\tilde\nu'_3)$. This formula has the same structure as the formula (\ref{eq25}).

\begin{figure}[htb]
\begin{center}
\includegraphics[width=0.7\linewidth]{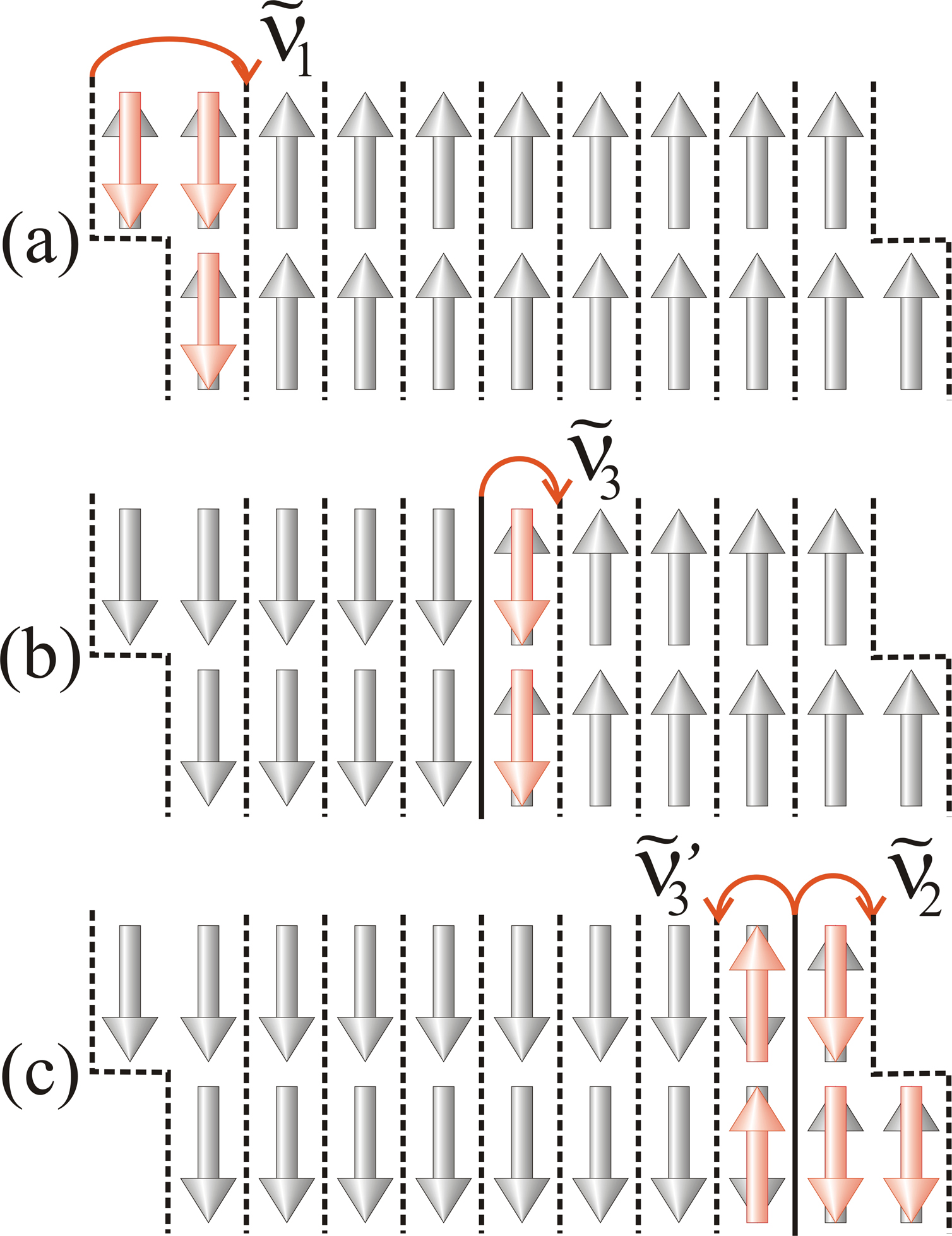}
\caption{\label{fig8} A schematic view of the biatomic chain of type B consisting of $2N=20$ magnetic moments. Dashed lines show the stable positions of the domain wall in a strong coupling approximation. Solid lines show its current position. The following processes are shown: (a) formation of the domain wall with the rate of $\tilde\nu_1$, (b) motion of the domain wall with the rate of $\tilde\nu_3$, and (c) the domain wall disappearance with the rate of $\tilde\nu_2$ and motion of the domain wall near the end of the chain with the rate of $\tilde\nu'_3$. The red arrows show the rotating magnetic moments.
}
\end{center}
\end{figure}

Now we consider the spontaneous remagnetization of the biatomic chain of type B. Figure~\ref{fig8} shows the positions of the domain wall corresponding to the local minima of energy. The rate $\tilde\nu_3$ of the domain wall motion obviously does not depend on the type of the biatomic chain. The value of $\tilde\nu_3$ is determined by the formula (\ref{eq16b}). The calculation of the rates $\tilde\nu_1$, $\tilde\nu_2$, and $\tilde\nu'_3$ is similar to the case A. Here, we present only the final formulas:
\begin{multline}\label{eq25b}
\tilde\nu_1=\nu''_1\left[\frac{\nu'_1\nu_3}{\nu'_1+\nu'_3}+\frac{\nu_1\nu'_3}{\nu'_2+\nu'_3}\right]\cdot\\
\cdot\left\{\nu_1+\nu''_1+\nu''_2+\nu_3+\frac{\nu_3(\nu''_1-\nu'_3)}{\nu'_1+\nu'_3}+
\frac{\nu_1(\nu''_1-\nu'_2)}{\nu'_2+\nu'_3}\right\}^{-1},
\end{multline}
\begin{equation}\label{eq26b}
\tilde\nu_2=\nu''_2\nu'_3\frac{\nu'_2\nu_3(\nu'_1+\nu'_3)+\nu_2\nu'_3(\nu'_2+\nu'_3)}{(\nu_3+\nu'_3)F_1+\nu'_3F_2},
\end{equation}
\begin{equation}\label{eq27b}
\tilde\nu'_3=\frac{\nu_3\nu'_3F_1}{(\nu_3+\nu'_3)F_1+\nu'_3F_2},
\end{equation}
where
\begin{multline}\label{eq28b}
F_1=(\nu_1+\nu''_2+\nu_3)(\nu'_1+\nu'_3)(\nu'_2+\nu'_3)-\\
-\nu_3\nu'_3(\nu'_2+\nu'_3)-\nu_1\nu'_2(\nu'_1+\nu'_3),
\end{multline}
\begin{multline}\label{eq29b}
F_2=(\nu_1+\nu''_2+\nu_3)[\nu_3(\nu'_1+\nu'_3)+\nu_2(\nu'_2+\nu'_3)]+\\
+(\nu'_2-\nu'_3)(\nu_3^2-\nu_1\nu_2)+\nu'_2\nu_3(\nu'_1+\nu'_3)+\nu_2\nu'_3(\nu'_2+\nu'_3).
\end{multline}
To estimate the reversal time of the magnetization $\tau^\text{strong}$ we need to substitute the rates (\ref{eq25b}), (\ref{eq26b}) and (\ref{eq27b}) into the formula (\ref{eq24b}) and to make the replacement $N\to N-1$.

\subsection{Interaction with STM}\label{STM_interaction}

\begin{figure}[htb]
\begin{center}
\includegraphics[width=0.95\linewidth]{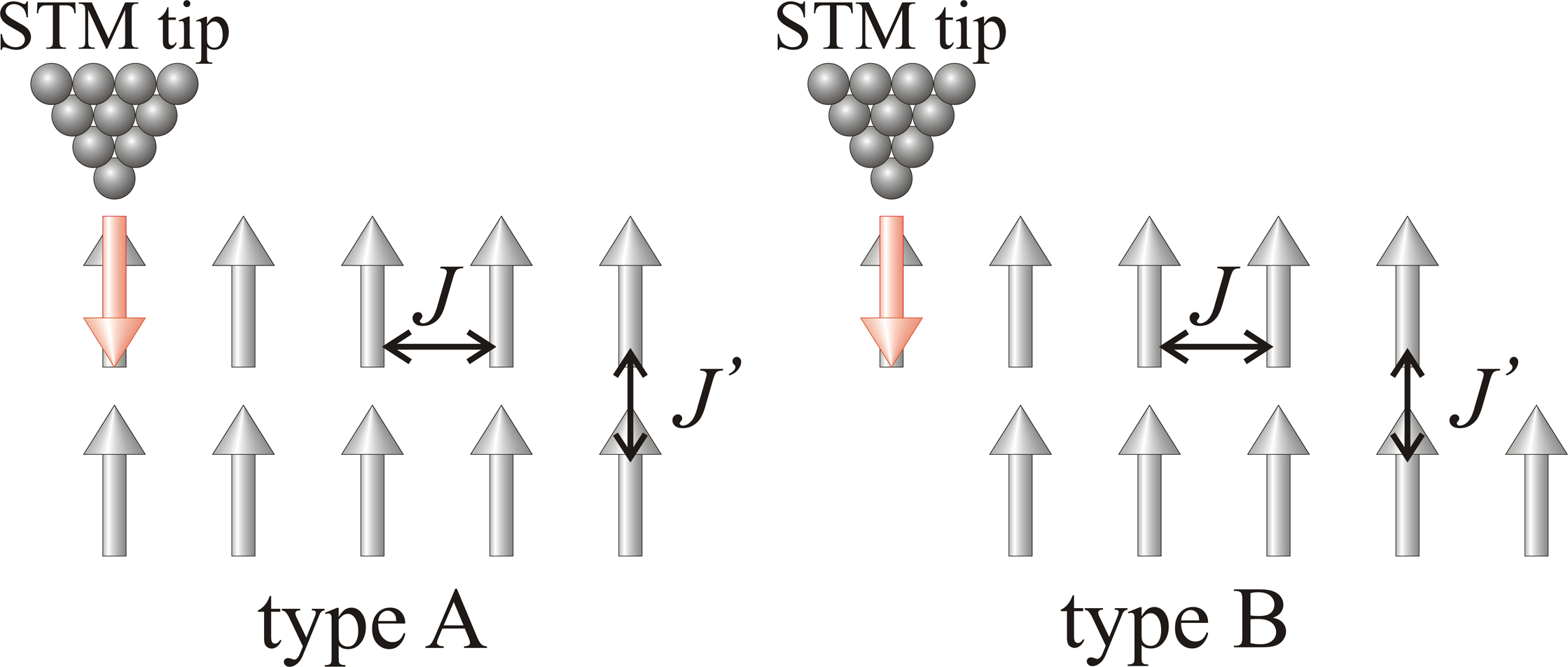}
\caption{\label{fig9} A schematic view of the interaction between the STM tip and the biatomic chains of types A and B. Red arrows show the magnetic moment which rotates as a result of this interaction.
}
\end{center}
\end{figure}

As shown in work~\cite{PRL110.087201}, the remagnetization of the antiferromagnetic chains occurs due to the formation of a domain wall at high voltages between the surface and the STM tip. In this Section we consider this regime of remagnetization. We assume that the atom located under the STM tip immediately flips and cannot return to its initial state. Then the reversal time of the magnetization of the biatomic chain $\tau_\text{STM}$ is the time of the reversal time of the magnetization of all other atoms. We assume that the STM tip is located as in the experimental work~\cite{science335.196}, i.e. over one of the edge atoms of the biatomic chain. The location of the STM tip and the magnetic moments of the atoms at the initial moment of time are shown in Fig.~\ref{fig9}. We work in a single domain-wall approximation and we assume that the domain wall is already formed at the initial moment of time near the STM tip. So, the remagnitization of the biatomic chain always occurs only from one end in the framework of our model. Further solution of the problem is different in cases of a  weak and a strong interaction between the atomic chains.

\subsubsection{A weak coupling approximation}

In the case of a weak coupling between the atomic chains we assume that in the beginning only the first chain which interacts with the STM tip is remagnetizing. The remagnetization of the second chain starts only when the remagnetization of the first chain is finished. Therefore, the total reversal time of the magnetization is equal to the sum of the reversal times of the magnetization of two atomic chains.
\begin{equation}\label{eq30b}
\tau^\text{weak}_\text{STM}=\tau_{+}^\text{STM}+\tau_{-},
\end{equation}
where the value of $\tau_{-}$ is calculated by the formula (\ref{eq7b}) or (\ref{eq11b}), for the chains of type A or B, respectively. The reversal time of the magnetization $\tau_{+}^\text{STM}$ does not depend on the type of the chain (see Fig.~\ref{fig9}). Its value can be determined by comparing the formulas (\ref{eq25}), (\ref{eq1a}), (\ref{eq2a}) and (\ref{eq38}), in which we need to make the replacement $\mu B\to|J'|$
\begin{multline}\label{eq31b}
\tau_{+}^\text{STM}=\frac{a_{-}}{\nu'_{3}(1-a_{-})}+\\
+\frac{(N-2)(1-a_{-})+(a_{-}-\alpha)S_{N-2}}{\nu_{3}(1-\alpha)(1-a_{-})},
\end{multline}
where $\alpha=(1-b)/b$, $S_N=(1-\alpha^N)/(1-\alpha)$, $a_{-}=\nu'_{3}/(\nu_{2}+\nu'_{3})$, and $b=\nu_{3}/(\nu_{3}+\nu'_{3})$.

\subsubsection{A strong coupling approximation}

In the case of a strong interaction between the chains, we should perform the calculations similar to those presented in Section~\ref{free_chain}. The rate of motion of the domain wall along the chain $\tilde\nu_3$ is still determined by the formula (\ref{eq16b}). The rates $\tilde\nu_1$, $\tilde\nu_2$, and $\tilde\nu'_3$ for the free end of the biatomic chain are also not changed. We have to calculate the rates $\tilde\nu_1^\text{STM}$, $\tilde\nu_2^\text{STM}$, and $\tilde{\nu'}_3^\text{STM}$ for the end of the biatomic chain interacting with the STM tip.  Then we will generalize the formula (\ref{eq24b}) to the case of the chains with non-equivalent ends.

The calculation of the rates is completely analogous to the one presented above. For the chains of type A we find
\begin{equation}\label{eq32b}
\tilde\nu_1^\text{STM}=\nu'_1,
\end{equation}
\begin{equation}\label{eq33b}
\tilde\nu_2^\text{STM}=\frac{\nu_2\nu'_3}{\nu_3+\nu'_3},
\end{equation}
\begin{equation}\label{eq34b}
\tilde{\nu'}_3^\text{STM}=\frac{\nu_3\nu'_3}{\nu_3+\nu'_3}.
\end{equation}

For the chains of type B we find
\begin{equation}\label{eq35b}
\tilde\nu_1^\text{STM}=\frac{\nu'_1\nu_3(\nu'_2+\nu'_3)+\nu_1\nu'_3(\nu'_1+\nu'_3)}
{(\nu'_1+\nu'_3)(\nu'_2+\nu'_3)+\nu_3(\nu'_2+\nu'_3)+\nu_1(\nu'_1+\nu'_3)},
\end{equation}
\begin{equation}\label{eq35b}
\tilde\nu_2^\text{STM}=\frac{\nu'_3}{F_3}\left[\nu'_2\nu_3(\nu'_1+\nu'_3)+\nu_2\nu'_3(\nu'_2+\nu'_3)\right],
\end{equation}
\begin{equation}\label{eq36b}
\tilde{\nu'}_3^\text{STM}=\frac{\nu_3\nu'_3}{F_3}(\nu'_1+\nu'_3)(\nu'_2+\nu'_3),
\end{equation}
\begin{multline}\label{eq37b}
F_3=(\nu'_1+\nu'_3)(\nu'_2+\nu'_3)(\nu_3+\nu'_3)+\\
+\nu_3\nu'_3(\nu'_1+\nu'_3)+\nu_2\nu'_3(\nu'_2+\nu'_3).
\end{multline}

Now the problem is reduced to the problem of finding the reversal time of the magnetization of the single-atomic chain. However, we cannot use the formula (\ref{eq24b}) because the rates of the magnetic moment flipping are different at the different ends of the biatomic chain ($\tilde\nu_1\ne\tilde\nu_1^\text{STM}$, $\tilde\nu_2\ne\tilde\nu_2^\text{STM}$, $\tilde\nu'_3\ne\tilde{\nu'}_3^\text{STM}$). The derivation of the formula for the reversal time of the magnetization of the single-atomic chain is similar to our previous calculations~\cite{Kolesnikov1, Kolesnikov2}. Below we discuss only the basic steps of the derivation. If the chain consists of $N$ atoms, then the domain wall can occupy $N + 1$ positions, as shown in Fig.~\ref{fig1}. The transition rates are equal to
\begin{equation}\label{eq38b}
\nu_{0\to1}=\tilde\nu_1^\text{STM},~~\nu_{1\to2}=\tilde{\nu'}_3^\text{STM},~~\nu_{N-1\to N}=\tilde\nu_2,
\end{equation}
\begin{multline}\label{eq39b}
\nu_{2\to3}=\dots=\nu_{N-2\to N-1}=\\
=\nu_{2\to1}=\dots=\nu_{N-2\to N-3}=\tilde\nu_3,
\end{multline}
\begin{equation}\label{eq40b}
\nu_{1\to0}=\tilde\nu_2^\text{STM},~~\nu_{N-1\to N-2}=\tilde\nu'_3,~~\nu_{N\to N-1}=\tilde\nu_1.
\end{equation}
According to the formula (\ref{eq15}) we find the transition probability matrix
\begin{equation}\label{eq41b}
T=
\begin{pmatrix}
0 & 1-a^\text{STM} & 0   & \ldots & 0 & 0& 0\\
1 & 0   & 1/2 & \ldots & 0 & 0& 0\\
0 & a^\text{STM}   & 0   & \ldots & 0 & 0& 0\\
\vdots&\vdots&\vdots&\ddots&\vdots&\vdots&\vdots\\
0 & 0&0&\ldots&0&1/2&0\\
0 & 0&0&\ldots&1/2&0&a\\
0 & 0&0&\ldots&0&1/2&0
\end{pmatrix},
\end{equation}
where $a=\tilde\nu'_3/(\tilde\nu_2+\tilde\nu'_3)$, $a^\text{STM}=\tilde{\nu'}_3^\text{STM}/(\tilde\nu_2^\text{STM}+\tilde{\nu'}_3^\text{STM})$. After solving the system of equations (\ref{eq18}) with $P^{\text{init}}_i=\delta_{0i}$, we find the reversal time of the magnetization by the formula (\ref{eq24})
\begin{multline}\label{eq42b}
\tau^\text{strong}_\text{STM}=\left[\frac{N-3}{2\tilde\nu_3}+\frac{1}{\tilde{\nu'}_3^\text{STM}}\right]\left[N-\frac{2(1-2a)}{1-a}\right]+\\
+\frac{a}{1-a}\left(\frac{1}{\tilde{\nu'}_3^\text{STM}}-\frac{1}{\tilde\nu'_3}\right)+\frac{1}{a^\text{STM}\tilde\nu_1^\text{STM}}\cdot\\
\left[N(1-a^\text{STM})-2(1-2a^\text{STM})+\frac{a-a^\text{STM}}{1-a}\right].
\end{multline}
Note that there is no factor $1/2$ in the formula (\ref{eq42b}) because we assume that the remagnetization always begins under the STM tip. This assumption is true if $\tau^\text{strong}_\text{STM}\ll\tau^\text{strong}$. We should make the replacement $N\to N-1$ in the formula (\ref{eq42b}) for the biatomic chain of type B.

\subsection{Ferromagnetic chains in the external magnetic field}\label{B_field}

In this Section, we consider the remagnetization of the biatomic chains in the external magnetic field ${\bf B}$ applied along the easy axis of magnetization ${\bf e}$. We focus on the most physically important case of the ferromagnetic chains ($ J, J '> 0 $). As can be seen from the works~\cite{PRL93.077203,NJPhys17.023014}, a strong coupling approximation $(J\approx J')$ is more applicable for the ferromagnetic chains. Here we consider only this approximation~\footnote{The case of a weak coupling $J'\ll J$ is much simpler. The estimations in this limit can be obtained by generalizing the formulas (\ref{eq5b}). The generalization of the formulas obtained below to the case of antiferromagnetic chains is also not difficult.}.

Following the work~\cite{Kolesnikov1}, we assume that the magnetic moments of all of the atoms are directed up at the initial moment of time, and ${\bf B}=B{\bf e}$, where $B$ can be both larger and less than zero. Instead of eight rates $\nu_1$, $\nu'_1$, $\nu''_1$, $\nu_2$, $\nu'_2$, $\nu''_2$, $\nu_3$, $\nu'_3$ used above, we will need the following sixteen rates: $\nu_{1\pm}$, $\nu'_{1\pm}$, $\nu''_{1\pm}$, $\nu_{2\pm}$, $\nu'_{2\pm}$, $\nu''_{2\pm}$, $\nu_{3\pm}$, $\nu'_{3\pm}$. Note that if the domain wall was in the position $i=0$ (see Fig.~\ref{fig1}) at the initial moment of time, and all magnetic moments were directed up, then the index ``$+$'' corresponds to motion of the domain wall to the right ($i\to i+1$), and the index ``$-$'' corresponds to motion of the domain wall to the left $i\to i-1$.

In a strong coupling approximation, we first need to calculate the rates $\tilde\nu_{1\pm}$, $\tilde\nu_{2\pm}$, $\tilde\nu_{3\pm}$, and $\tilde\nu'_{3\pm}$. Let us calculate the rates $\tilde\nu_{3\pm}$ which do not depend on the type of the biatomic chain. As before, the domain wall can transit from the initial state 0 to the final states I or II through the transient states 1,2,3,4 (see Fig.~\ref{fig5}). However, now the rates of motion of the domain wall to the right and left are different from each other. Instead of formulas (\ref{eq14b}) and (\ref{eq15b}) we find
\begin{equation}\label{eq43b}
\nu_{0\to 1}=\nu_{0\to 3}=\nu_{3+},~~~~\nu_{0\to 2}=\nu_{0\to 4}=\nu_{3-},
\end{equation}
\begin{equation}\label{eq44b}
\nu_{2\to 0}=\nu_{4\to 0}=\nu_{1\to I}=\nu_{3\to I}=\nu'_{3+},
\end{equation}
\begin{equation}\label{eq45b}
\nu_{1\to 0}=\nu_{3\to 0}=\nu_{2\to II}=\nu_{4\to II}=\nu'_{3-}.
\end{equation}
The non-zero elements of the transition probability matrix calculated by the formula (\ref{eq15}) are equal to $T_{01}=T_{03}=1-b'$, $T_{02}=T_{04}=b'$, $T_{10}=T_{30}=b/2$, $T_{20}=T_{40}=(1-b)/2$, where $b=\nu_{3+}/(\nu_{3+}+\nu_{3-})$ and $b'=\nu'_{3+}/(\nu'_{3+}+\nu'_{3-})$. After solving the system of equations (\ref{eq18}) with $P^{\text{init}}=\{1,0,0,0,0\}$, we find
\begin{multline}\label{eq46b}
\tilde\nu_{3+}=\frac{\tau_1^1P_1\nu_{1\to I}+\tau_3^1P_3\nu_{3\to I}}{\tau_\text{tot}}=\\
=\frac{2\nu_{3+}\nu'_{3+}}{(\nu'_{3+}+\nu'_{3-})+2(\nu_{3+}+\nu_{3-})},
\end{multline}
\begin{multline}\label{eq47b}
\tilde\nu_{3-}=\frac{\tau_2^1P_2\nu_{2\to II}+\tau_4^1P_4\nu_{4\to II}}{\tau_\text{tot}}=\\
=\frac{2\nu_{3-}\nu'_{3-}}{(\nu'_{3+}+\nu'_{3-})+2(\nu_{3+}+\nu_{3-})}.
\end{multline}
We note the following.  First,  if $B\to-B$, then the formula (\ref{eq46b}) turns to (\ref{eq47b}). Second, if $B\to0$, then the formulas (\ref{eq46b}) and (\ref{eq47b}) tend to (\ref{eq16b}).

The rates $\tilde\nu_{1\pm}$, $\tilde\nu_{2\pm}$, and $\tilde\nu'_{3\pm}$ can be calculated in the same way. For the biatomic chain of type A we find
\begin{equation}\label{eq48b}
\tilde\nu_{1\pm}=\frac{2\nu_{1\pm}\nu'_{1\pm}}{\nu_{1\pm}+\nu'_{1\pm}+\nu'_{2\mp}},
\end{equation}
\begin{multline}\label{eq49b}
\tilde\nu_{2\pm}=\\
=\frac{2\nu_{2\pm}\nu'_{2\pm}(\nu'_{3\pm}+\nu'_{3\mp})}
{2\nu_{3\mp}(\nu'_{1\mp}+\nu'_{2\pm})+(\nu'_{3\pm}+\nu'_{3\mp})(2\nu_{2\pm}+\nu'_{1\mp}+\nu'_{2\pm})},
\end{multline}
\begin{multline}\label{eq50b}
\tilde\nu'_{3\pm}=\\
=\frac{2\nu_{3\pm}\nu'_{3\pm}(\nu'_{1\pm}+\nu'_{2\mp})}
{2\nu_{3\pm}(\nu'_{1\pm}+\nu'_{2\mp})+(\nu'_{3\pm}+\nu'_{3\mp})(2\nu_{2\mp}+\nu'_{1\pm}+\nu'_{2\mp})}.
\end{multline}
For the biatomic chain of type B we find
\begin{multline}\label{eq51b}
\tilde\nu_{1\pm}=\nu''_{1\pm}\left[\frac{\nu'_{1\pm}\nu_{3\pm}}{\nu'_{1\pm}+\nu'_{3\mp}}+\frac{\nu_{1\pm}\nu'_{3\pm}}{\nu'_{2\mp}+\nu'_{3\pm}}\right]\cdot\\
\cdot\left\{\nu_{1\pm}+\nu''_{1\pm}+\nu''_{2\mp}+\nu_{3\pm}+\frac{\nu_{3\pm}(\nu''_{1\pm}-\nu'_{3\mp})}{\nu'_{1\pm}+\nu'_{3\mp}}+\right.\\
+\left.\frac{\nu_{1\pm}(\nu''_{1\pm}-\nu'_{2\mp})}{\nu'_{2\mp}+\nu'_{3\pm}}\right\}^{-1},
\end{multline}
\begin{multline}\label{eq52b}
\tilde\nu_{2\pm}=\nu''_{2\pm}(\nu'_{3\pm}+\nu'_{3\mp})\cdot\\
\frac{\nu'_{2\pm}\nu_{3\pm}(\nu'_{1\mp}+\nu'_{3\pm})+\nu_{2\pm}\nu'_{3\pm}(\nu'_{2\pm}+\nu'_{3\mp})}
{(2\nu_{3\mp}+\nu'_{3\pm}+\nu'_{3\mp})F_{1\mp}+(\nu'_{3\pm}+\nu'_{3\mp})F_{2\mp}},
\end{multline}
\begin{equation}\label{eq53b}
\tilde\nu'_{3\pm}=\frac{2\nu_{3\pm}\nu'_{3\pm}F_{1\pm}}{(2\nu_{3\pm}+\nu'_{3\pm}+\nu'_{3\mp})F_{1\pm}+(\nu'_{3\pm}+\nu'_{3\mp})F_{2\pm}},
\end{equation}
where
\begin{multline}\label{eq54b}
F_{1\pm}=(\nu_{1\pm}+\nu''_{2\mp}+\nu_{3\pm})(\nu'_{1\pm}+\nu'_{3\mp})(\nu'_{2\mp}+\nu'_{3\pm})-\\
-\nu_{3\pm}\nu'_{3\mp}(\nu'_{2\mp}+\nu'_{3\pm})-\nu_{1\pm}\nu'_{2\mp}(\nu'_{1\pm}+\nu'_{3\mp}),
\end{multline}
\begin{multline}\label{eq55b}
F_{2\pm}=(\nu'_{2\mp}-\nu'_{3\mp})(\nu_{3\pm}\nu_{3\mp}-\nu_{1\pm}\nu_{2\mp})+\\
+(\nu_{1\pm}+\nu''_{2\mp}+\nu_{3\pm})[\nu_{3\mp}(\nu'_{1\pm}+\nu'_{3\mp})+\nu_{2\mp}(\nu'_{2\mp}+\nu'_{3\pm})]+\\
+\nu'_{2\mp}\nu_{3\mp}(\nu'_{1\pm}+\nu'_{3\mp})+\nu_{2\mp}\nu'_{3\mp}(\nu'_{2\mp}+\nu'_{3\pm}).
\end{multline}

Using the found rates, we calculate the reversal time of the magnetization of the biatomic ferromagnetic chain in the external magnetic field. Now the problem is reduced to the estimation of the reversal time of the magnetization of the single-atomic chain shown in Fig.~\ref{fig1}. The transition rates are the following
\begin{equation}\label{eq56b}
\nu_{0\to1}=\tilde\nu_{1+},~~~\nu_{1\to2}=\tilde{\nu'}_{3+},~~~\nu_{N-1\to N}=\tilde\nu_{2+},
\end{equation}
\begin{equation}\label{eq57b}
\nu_{2\to3}=\dots=\nu_{N-2\to N-1}=\tilde\nu_{3+},
\end{equation}
\begin{equation}\label{eq58b}
\nu_{1\to0}=\tilde\nu_{2-},~~\nu_{N-1\to N-2}=\tilde\nu'_{3-},~~\nu_{N\to N-1}=\tilde\nu_{1-},
\end{equation}
\begin{equation}\label{eq59b}
\nu_{2\to1}=\dots=\nu_{N-2\to N-3}=\tilde\nu_{3-}.
\end{equation}
According to the formula (\ref{eq15}) we find the transition probability matrix
\begin{equation}\label{eq60b}
T=
\begin{pmatrix}
0 & 1-a'_{+} & 0   & \ldots & 0 & 0& 0\\
1 & 0   & 1-b & \ldots & 0 & 0& 0\\
0 & a'_{+}   & 0   & \ldots & 0 & 0& 0\\
\vdots&\vdots&\vdots&\ddots&\vdots&\vdots&\vdots\\
0 & 0&0&\ldots&0&1-b&0\\
0 & 0&0&\ldots&b&0&a'_{-}\\
0 & 0&0&\ldots&0&b&0
\end{pmatrix},
\end{equation}
where $a'_{+}=\tilde\nu'_{3+}/(\tilde\nu_{2-}+\tilde\nu'_{3+})$, $a'_{-}=\tilde\nu'_{3-}/(\tilde\nu_{2+}+\tilde\nu'_{3-})$ and $b=\tilde\nu_{3+}/(\tilde\nu_{3-}+\tilde\nu_{3+})$. After solving the system of equations (\ref{eq18}) with $P^{\text{init}}_i=\delta_{0i}$, we find the reversal time of the magnetization by the formula (\ref{eq24})
\begin{multline}\label{eq61b}
\tau^\text{strong}_\text{B}(B)=
=\frac{1}{2(1-a'_{-})}\left\{\frac{a'_{-}}{\tilde\nu'_{3-}}+\right.\\
+\frac{(N-2)(1-a'_{-})+(a'_{-}-\alpha)S_{N-2}}{\tilde\nu_{3+}(1-\alpha)}+\\
+\frac{S_{N-2}-(a'_{-}+\alpha a'_{+})S_{N-3}+\alpha a'_{+}a'_{-}S_{N-4}}{\tilde\nu_{1+}a'_{+}}+\\
+\left.\left(\frac{1}{\tilde\nu'_{3+}}-\frac{1}{\tilde\nu_{3+}}\right)(1-\alpha)\left[1-(a'_{-}-\alpha)S_{N-3}\right]\right\},
\end{multline}
where $\alpha=(1-b)/b$, $S_N=(1-\alpha^N)/(1-\alpha)$. The formula (\ref{eq61b}) is valid for the biatomic chains of type A. It is necessary to make the replacement of $N\to N-1$ for the chains of type B. The formula (\ref{eq61b}) has the same structure as the formula (\ref{eq38}) and tends to the formula (\ref{eq24b}) in the limit of $B\to0$.

The formula (\ref{eq61b}) can be used to study the magnetodynamic properties of the biatomic chains at the temperatures below $T_{\text{max}}$. If the magnetic field $B$ is a function of time $B=B(t)$, then the rates of remagnetization of the biatomic chains $\nu_{\uparrow\to\downarrow}(t)=1/\tau^\text{strong}_\text{B}(B(t))$ and $\nu_{\downarrow\to\uparrow}(t)=1/\tau^\text{strong}_\text{B}(-B(t))$ are also functions of time. The probability of finding the biatomic chain in the state when the magnetic moments of all of the atoms are directed up can be find from the master equation
\begin{equation}\label{eq39}
\frac{dP_{\uparrow}}{dt}=P_{\downarrow}\nu_{\downarrow\to\uparrow}-P_{\uparrow}\nu_{\uparrow\to\downarrow},
\end{equation}
where $P_{\uparrow}+P_{\downarrow}=1$~\footnote{We assume that the average time of random walk of the domain wall is much less than the average time of the formation of new domain wall.This condition is satisfied if $T<T_{\text{max}}$~\cite{Kolesnikov1}.}. If the magnetization of the biatomic chain is measured in arbitrary units $M\in[-1,1]$, then $M=P_{\uparrow}-P_{\downarrow}$. And we find from the master equation (\ref{eq39}) the following equation for the magnetization of the biatomic chain
\begin{equation}\label{eq40}
\frac{dM(t)}{dt}=\mathfrak{A}(t)M(t)+\mathfrak{B}(t),
\end{equation}
where $\mathfrak{A}=-\nu_{\uparrow\to\downarrow}-\nu_{\downarrow\to\uparrow}$ and $\mathfrak{B}=\nu_{\downarrow\to\uparrow}-\nu_{\uparrow\to\downarrow}$. The equation (\ref{eq40}) together with the initial condition $M(0)=M_0$ is the Cauchy problem, which can be easily solved numerically.

\subsection{Numerical estimates}\label{Numerical}

In order to demonstrate the applicability of our method, let us consider the numerical estimates for two physical systems. The first system is the antiferromagnetic Fe chains  on Cu$_2$N/Cu(001) surface. According to the experimental work~\cite{science335.196}, the exchange energies of Fe atoms are $J=1.3\pm0.1$~meV and $J'=0.03\pm0.02$~meV. In three-atomic chain MAE varies from $2.1\pm0.1$~meV for the edge atoms to $3.6\pm0.1$~meV for the central atom~\cite{NatureNano10.40}. For the numerical estimates we choose the following parameters of the Hamiltonian $J=1.3$~meV, $J'=0.03$~meV, $K=3$~meV. We consider short chains consisting of $2N=20$ atoms. Note that with this choice of parameters $J/J'\approx43$ and $J/(NJ')\approx4.3$. Thus, a weak coupling approximation should work well. The critical temperature $T_\text{C}$ for a single-atomic chain is estimated by means of the kMC method~\cite{PhysRevB.73.174418}. We found that $T_\text{C}$ decreases monotonically with increasing of the chain length from $10\pm1$~K at $N=10$ to $6\pm1$~K at $N=100$. Obviously, the critical temperature of the biatomic chain is higher than $T_\text{C}$ of the single-atomic chain. The most of numerical estimations will be performed at the temperature $T=4$~K. The Fe biatomic chains are definitely in the antiferromagnetic state at this temperature. By varying the parameters of the Hamiltonian~(\ref{eq2}), we will find the applicability limits of our method.

\begin{figure}[htb]
\begin{center}
\includegraphics[width=0.95\linewidth]{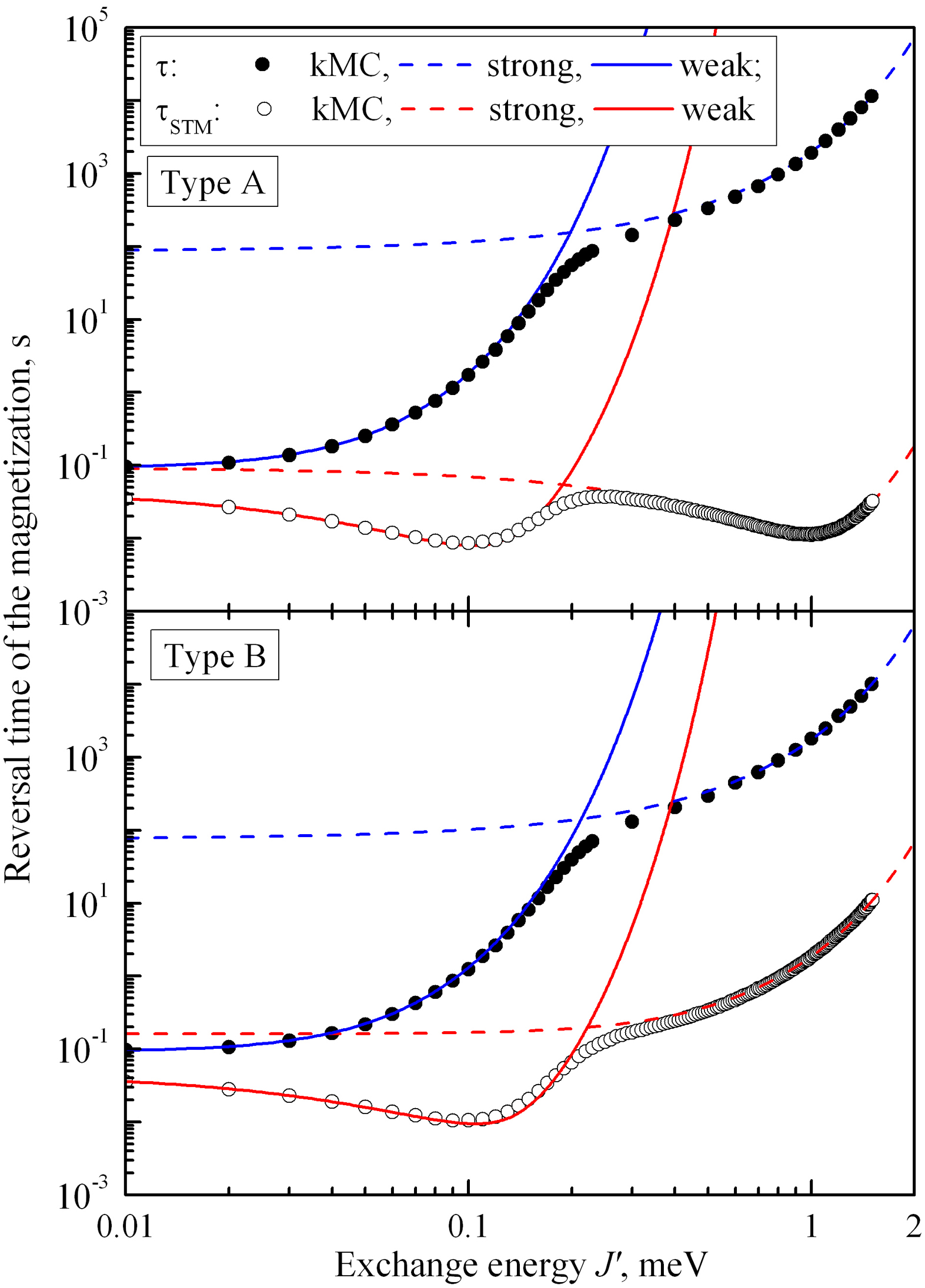}
\caption{\label{fig10} Dependencies of the reversal time of the magnetization on the exchange energy $J'\in[0.01,1.5]$~meV. The other parameters of Heisenberg Hamiltonian are the following: $J=1.3$~meV, $K=3$~meV, $T=4$~K, $2N=20$. The reversal times of the magnetization averaged over 10000 kMC simulations are shown with points. Solid (dashed) lines correspond to the approximation of a weak (strong) coupling between the chains.}
\end{center}
\end{figure}

Figure~\ref{fig10} shows the dependencies of the reversal time of the magnetization in the cases of the spontaneous remagnetization $\tau$ and the remagnetization under the interaction with the STM tip $\tau_\text{STM}$. Solid and dashed lines correspond to a weak and a strong approximation, respectively. The points show the results of the kMC simulations~\footnote{In this and the following plots, the errors of the kMC simulation results do not exceed the size of the marker.}. The upper and the lower plots correspond to the biatomic chains of types A and B, respectively. The estimates of $\tau^\text{weak}$ and $\tau^\text{weak}_\text{STM}$ are in excellent agreement with the results of the kMC simulation at low exchange energies $J'\ll J$. A weak coupling approximation works well up to the value of $J'\approx J/N=0.13$~meV. A strong coupling approximation works well at $J'\approx J$. Figure~\ref{fig10} shows that a strong coupling approximation remains valid as $J'$ decreases down to the value of $J/\ln N=0.56$~meV. Both of the approximations are not very accurate in the intermediate range of $J/N\lesssim J'\lesssim J/\ln N$. However, the function $\min(\tau^\text{weak}_\text{(STM)},\tau^\text{strong}_\text{(STM)})$ can be used as the upper limits of the reversal time of the magnetization $\tau_\text{(STM)}$. Note that the reversal time of the magnetization in the case of the spontaneous remagnetization is a monotonically increasing function of $J'$ and it is slightly different for chains of types A and B. At the same time, the reversal time of the magnetization $\tau_\text{STM}$ is a non-monotonic function. The functions $\tau_\text{STM}(J')$ differ significantly from each other, especially at large $J'$. In the case B, the dependence $\tau_\text{STM}(J')$ has one local minimum. In the case A, it has two local minima and one local maximum.
It is important to note that the function $\min(\tau^\text{weak}_\text{(STM)},\tau^\text{strong}_\text{(STM)})$ qualitatively describes the behavior of the function $\tau_\text{(STM)}(J')$ at all values of $J'$~\footnote{We also tried to describe the dependence of $\tau_{(\text{STM})}(J')$ by the function  $\left[(\tau^\text{weak}_{(\text{STM})})^{-1}+(\tau^\text{strong}_{(\text{STM})})^{-1}\right]^{-1}$. Unfortunately, this function is not much better than the function $\min(\tau^\text{weak}_{(\text{STM})},\tau^\text{strong}_{(\text{STM})})$ and also does not lead to quantitative agreement with the results of the kMC simulations in the whole range of parameters.}.

\begin{figure}[htb]
\begin{center}
\includegraphics[width=0.95\linewidth]{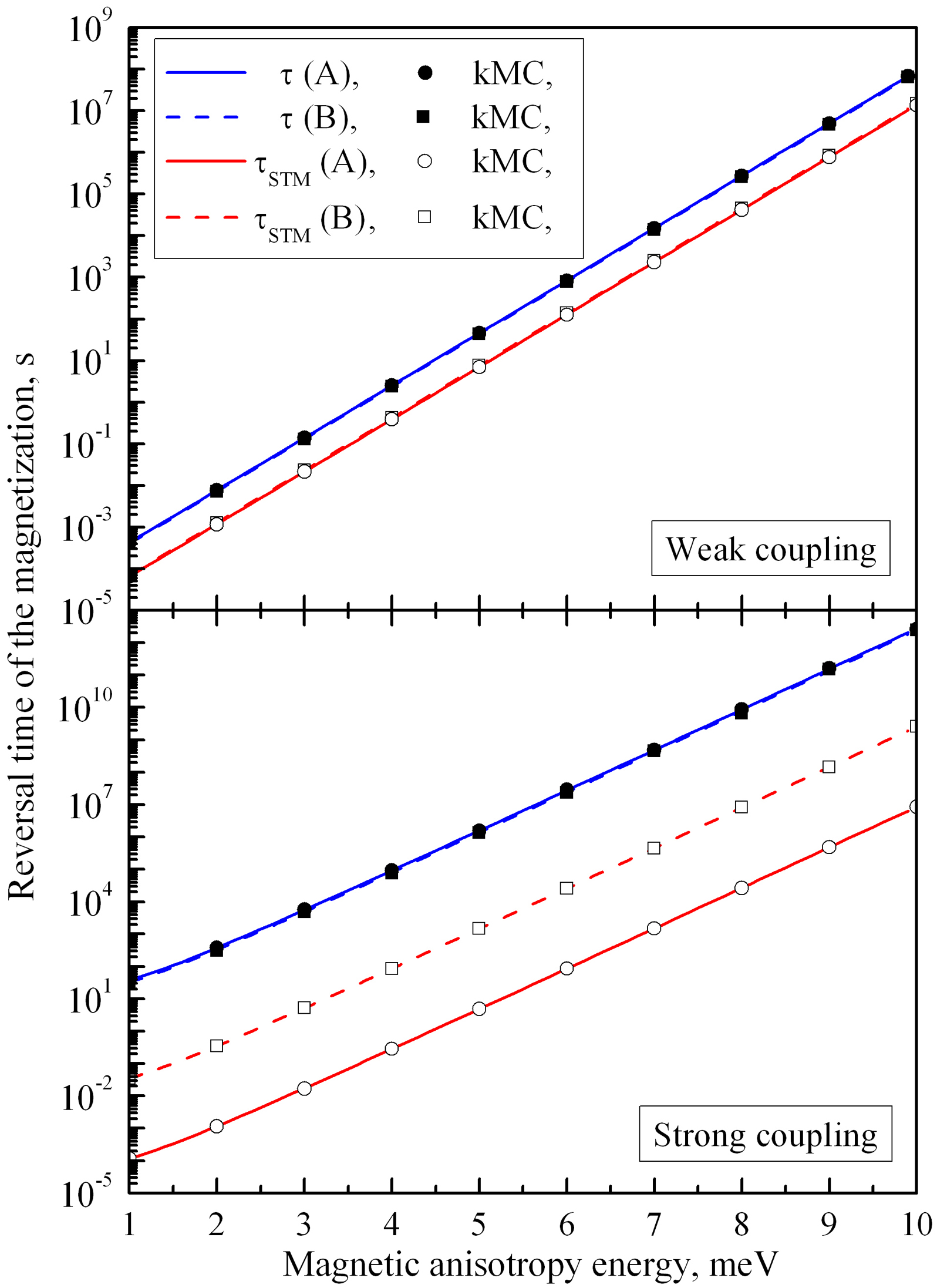}
\caption{\label{fig11} Dependencies of the reversal time of the magnetization on the MAE $K\in[1,10]$~meV. The other parameters of Heisenberg Hamiltonian are the following: $J=1.3$~meV, $J'=0.03$~meV (upper plot), $J'=1.3$~meV (lower plot), $T=4$~K, $2N=20$. The reversal times of the magnetization averaged over 10000 kMC simulations are shown with points. Solid and dashed lines correspond to the approximation of a weak (upper plot) and a strong (lower plot) coupling between the chains.
}
\end{center}
\end{figure}

The dependence of the reversal times of the magnetization $\tau$ and $\tau_\text{STM}$ on the MAE in the range of $K\in[1,10]$~meV is shown in Fig.~\ref {fig11}. The upper plot corresponds to the case of a weak coupling between the atomic chains $J=1.3$~meV, $J'=0.03$~meV, and the lower plot corresponds to the case of a strong coupling $J=J'=1.3$~meV. These dependencies have a very simple form $\ln\tau_{(\text{STM})}\sim K$. The reversal times of the magnetization of the biatomic chains of different types are similar in the case of a weak coupling between the atomic chains. But the reversal times of the magnetization $\tau_\text{STM}$ differ by 3 orders of magnitude in the case of a strong coupling between the atomic chains. As can be seen from the Fig.~\ref {fig11}, the reversal times of the magnetization calculated by analytical formulas perfectly agree with the results of kMC simulations. Note that the conditions of applicability of the single-domain approximation $K(2N)-2(J+J')\gtrsim k_\text{B}T$ (for the chain of type A) and $K(2N)-2\max(J,J')\gtrsim k_\text{B}T$ (for the chain of type B) are satisfied at $K\ge1$~meV.

\begin{figure}[htb]
\begin{center}
\includegraphics[width=0.95\linewidth]{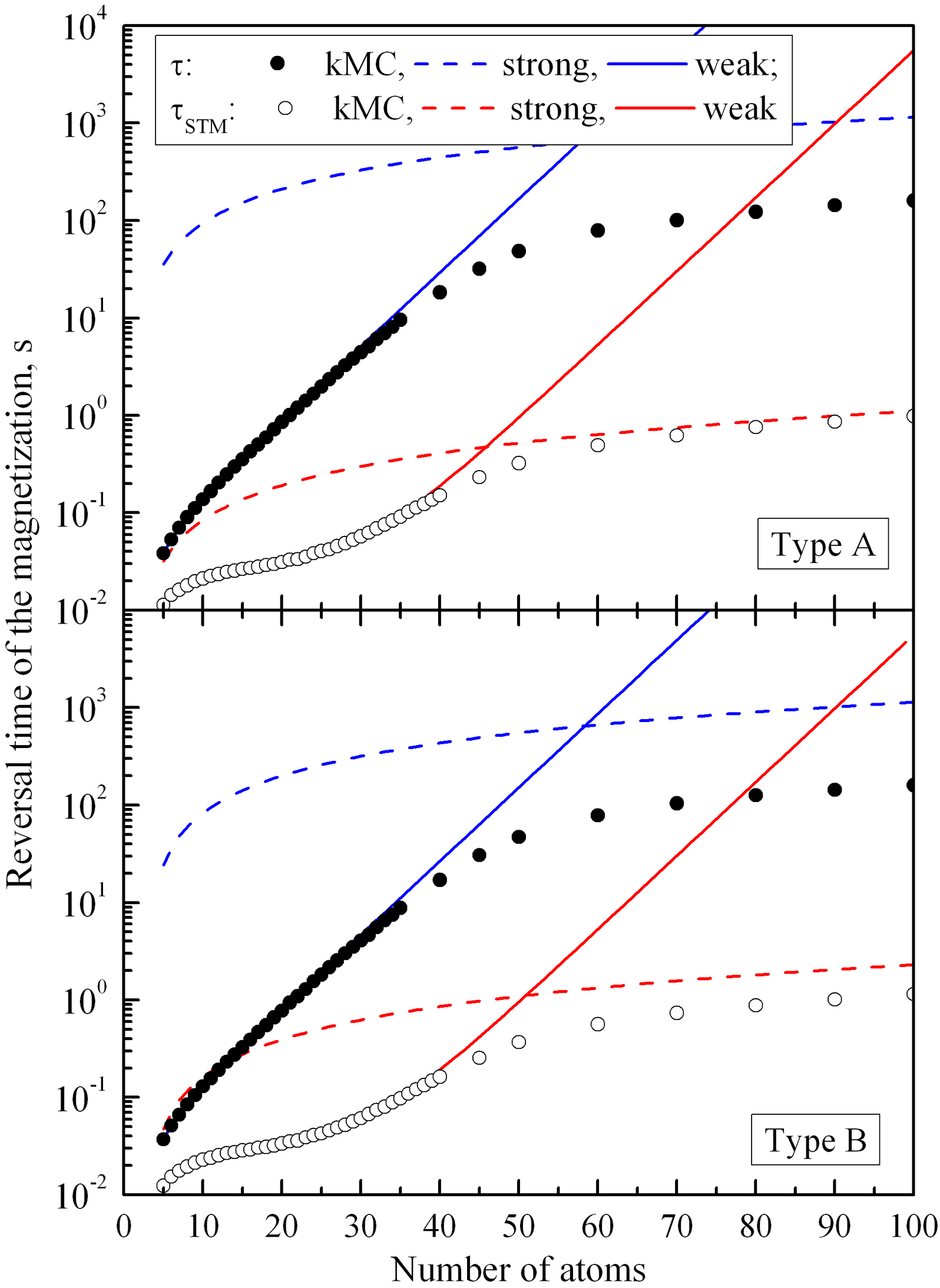}
\caption{\label{fig12} Dependencies of the reversal time of the magnetization on the number of atoms in each of the atomic chains $N\in[5,100]$. The total number of atoms in the biatomic chain is $2N$. The other parameters of Heisenberg Hamiltonian are the following: $J=1.3$~meV, $J'=0.03$~meV, $K=3$~meV, $T=4$~K. The reversal times of the magnetization averaged over 10000 kMC simulations are shown with points. Solid (dashed) lines correspond to the approximation of a weak (strong) coupling between the chains.
}
\end{center}
\end{figure}

The dependencies of the reversal times of the magnetization $\tau$ and $\tau_\text{STM}$ on the length of the biatomic chain $N$ are shown in Fig.~\ref{fig12}. The total number of atoms in the biatomic chain is $2N$. The upper and the lower plots correspond to biatomic chains of types A and B, respectively. The results obtained in the framework of a weak coupling approximation agrees well with the results of the kMC simulations at $N<J/J'\approx43$. With a further increase of $N$, a weak coupling approximation leads to a high overestimation of the reversal times of the magnetization. Indeed, $\tau^\text{weak}_{(\text{STM})}\sim e^N$ when $N>J/J'$, while the kMC simulations lead to a linear relationship $\tau_{(\text{STM})}\sim N$. A strong coupling approximation also does not work in this range of the parameters because the condition $J'\ln N\gtrsim J$ is obviously not satisfied. However, $\tau^\text{strong}_{(\text{STM})}\sim N\sim\tau_{(\text{STM})}$. Thus, the estimation $\tau^\text{strong}_{(\text{STM})}$ can be used as the upper limit on the value of $\tau_{(\text{STM})}$. For example, the estimate of $\tau^\text{strong}_\text{STM}$ is in a good agreement with the results of the kMC simulations already at $N\approx100$ in the case of the biatomic chain of type A.

\begin{figure}[htb]
\begin{center}
\includegraphics[width=0.95\linewidth]{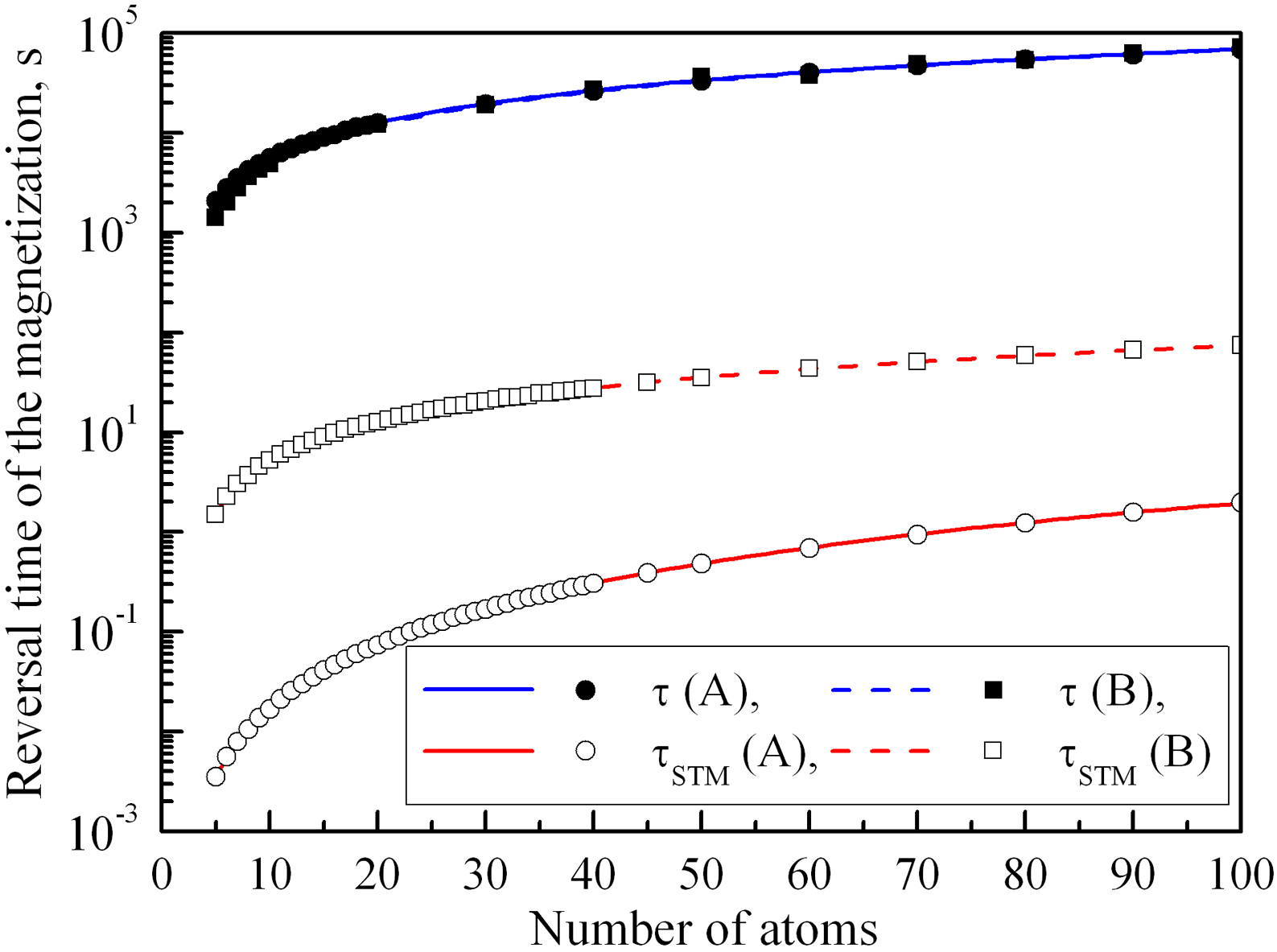}
\caption{\label{fig13} Dependencies of the reversal time of the magnetization on the number of atoms in each of the atomic chains $N\in[5,100]$. The total number of atoms in the biatomic chain is $2N$. The other parameters of Heisenberg Hamiltonian are the following: $J=J'=1.3$~meV, $K=3$~meV, $T=4$~K. The reversal times of the magnetization averaged over 10000 kMC simulations are shown with points. Solid and dashed lines correspond to the approximation of a strong coupling between the chains.
}
\end{center}
\end{figure}

Let us consider a hypothetical biatomic chain with a strong coupling between the atomic chains $ J = J '= 1.3 $ ~ meV. The dependencies of the reversal times of the magnetization $\tau$ and $\tau_\text{STM}$ on the length of the chain $N$ are presented in Fig.~\ref{fig13}. The dependencies $\tau_{(\text{STM})}(N)$ are close to the linear $\tau_{(\text{STM})}\sim N$ for $N\sim100$. The times $\tau$ are almost the same for the biatomic chains of types A and B. At the same time, the values of $\tau_\text{STM}$ in the case of the chains of type A are several orders lower than those in the case of the chains of type B. This is due to the fact that the atom interacting with the STM tip is more strongly coupled with the chain in the case A than in the case B. The reversal times of the magnetization calculated in a strong coupling approximation are in excellent agreement with the results of the kMC simulations.

\begin{figure}[htb]
\begin{center}
\includegraphics[width=0.95\linewidth]{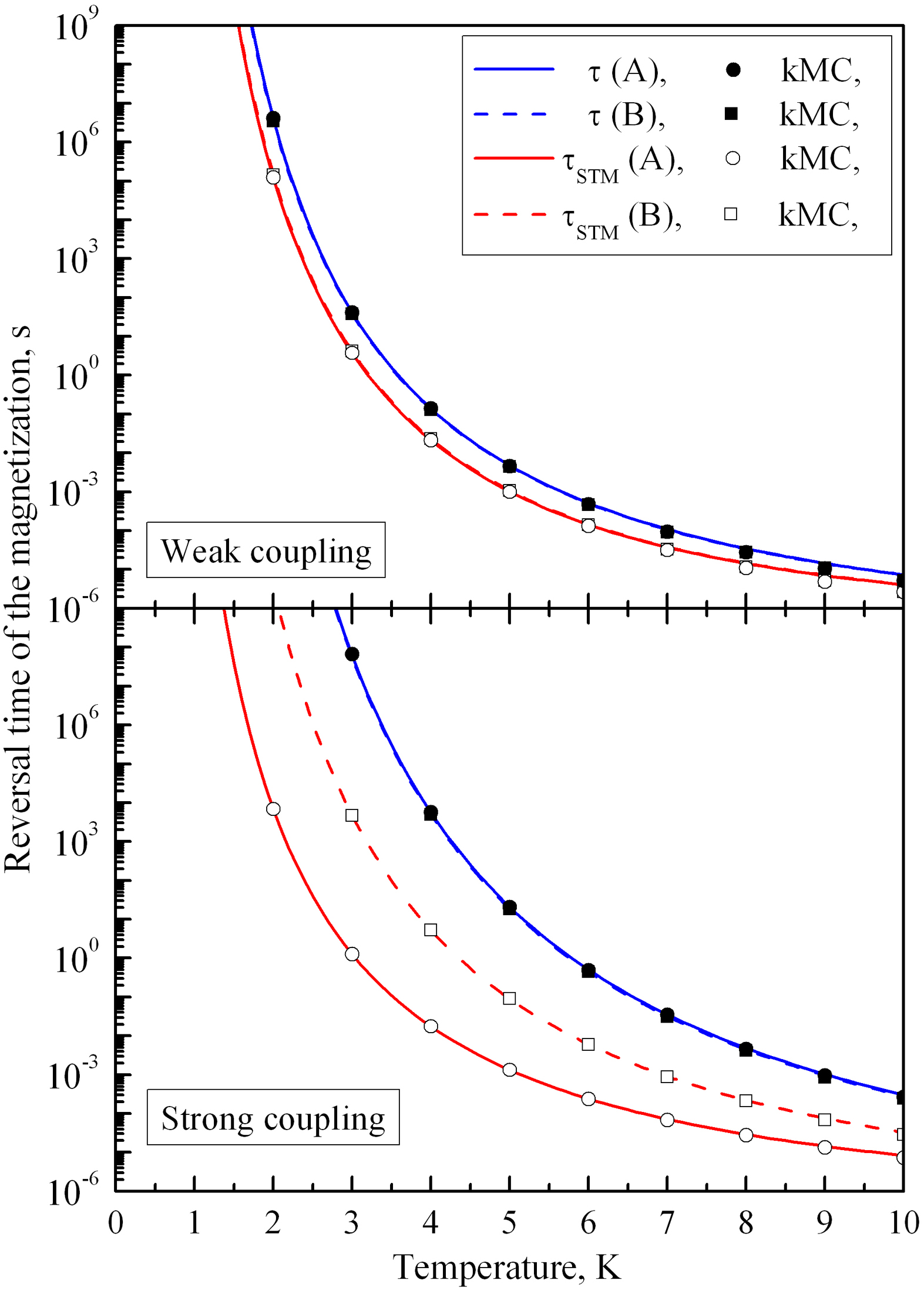}
\caption{\label{fig14} Dependencies of the reversal time of the magnetization on the temperature $T\in[0,10]$~K. The other parameters of Heisenberg Hamiltonian are the following: $J=1.3$~meV, $J'=0.03$~meV (upper plot), $J'=1.3$~meV (lower plot), $K=3$~meV, $2N=20$. The reversal times of the magnetization averaged over 10000 kMC simulations are shown with points. Solid and dashed lines correspond to the approximation of a weak (upper plot) and strong a (lower plot) coupling between the chains.
}
\end{center}
\end{figure}

Figure~\ref{fig14} shows the temperature dependence of the reversal times of the magnetization $\tau$ and $\tau_\text{STM}$. The upper plot corresponds to the biatomic chain with a weak coupling between the atomic chains $J/J'\approx43$. The lower plot corresponds to a hypothetical biatomic chain with a strong coupling $J'=J$. We see that in the both cases the values of $\tau_{(\text{STM})}$ calculated by analytical formulas are in excellent agreement with the results of the kMC simulations. Here it is necessary to make two important notes. First, let us discuss the value of the maximum temperature $T_\text{max}$ till which a single-domain approximation remains valid. The average time of formation of the domain wall $\tau_{+}$ and the average time of random walk of the domain wall $\tau_{walk}$ can be estimated~\cite{Kolesnikov1}. The temperature $T_\text{max}$ can be found as a solution of the equation $\tau_{+}=\tau_\text{walk}$. This equation can be written in the form (\ref{eq31}). However, in the case of short atomic chains $N\le100$ the simultaneous appearance of two or more domain walls actually means the transition to the paramagnetic phase. Thus, the proposed method for estimating of the reversal time of the magnetization of the biatomic chains leads to adequate results almost up to the critical temperature $T_\text{min}\approx T_\text{C}$. Second, the calculation time needed for the kMC simulations grows exponentially with the decrease of temperature. Therefore, the calculation of the values of $\tau_\text{STM}$ and $\tau$ at low temperatures by means of the kMC method is almost impossible. In this case, the proposed method seems to be the only possible opportunity for estimating of the reversal time of the magnetization.

The second system under the consideration is Co biatomic chains on Pt (997) surface. According to the works~\cite{Gambardella.Nature,PRL93.077203} the exchange energies are $J\approx J'\approx7.5$~meV, the MAE is $K=0.33\pm0.04$~meV for the biatomic Co chain. The magnetic moment of Co atom $\mu$ is the sum of the spin magnetic moment $\mu_S\approx2.08\mu_\text{B}$ and the orbital magnetic moment $\mu_L\approx0.37\mu_\text{B}$, where $\mu_\text{B}$ is the Bohr magneton. For the numerical estimates, we choose the following parameters of the Hamiltonian: $J=J'=7.5$~meV, $K=0.34$~meV, $\mu=2.4\mu_\text{B}$. In order to prevent the simultaneous flipping of the magnetic moments of the atoms (the superparamagnetic regime) the following inequalities must be satisfied: $KN-2J\gtrsim k_\text{B}T$ for the single-atomic chain, $K(2N)-2(J+J')\gtrsim k_\text{B}T$ for the biatomic chain of type A, and $K(2N)-2\max(J,J')\gtrsim k_\text{B}T$ for the biatomic chain of type B. For the single-atomic chains at the temperature of $T=30-70$~K, this condition is satisfied for chains longer than 70 atoms. Further, we  consider the biatomic chains consisting of $2N=200$ atoms, which are definitely ferromagnetic. We will see later that the critical temperature of such biatomic chains is approximately 70~K.

\begin{figure}[htb]
\begin{center}
\includegraphics[width=0.95\linewidth]{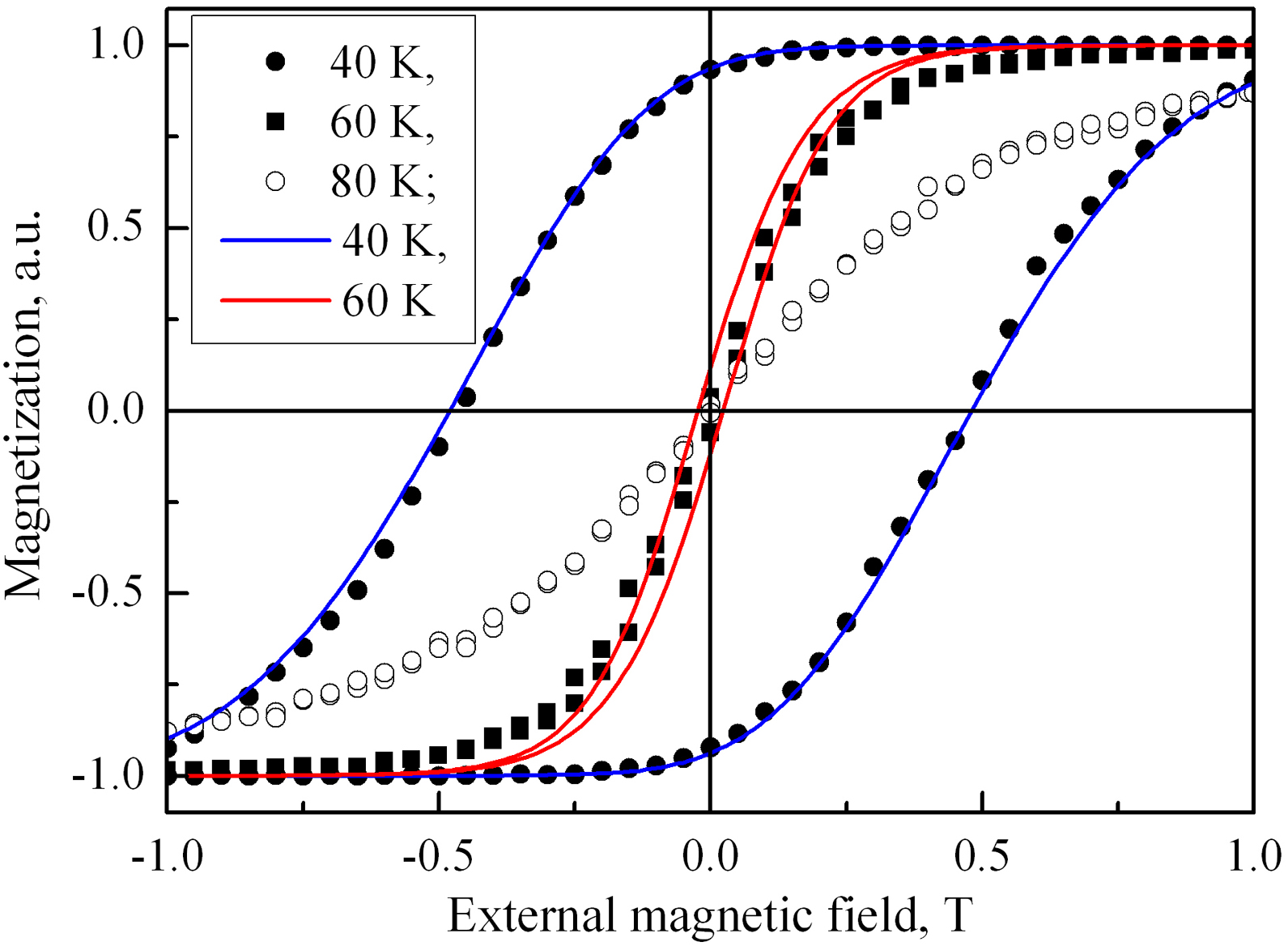}
\caption{\label{fig15} Magnetization response to the external magnetic field for the biatomic chain of type A consisting of $2N=200$ atoms at three temperatures: 40 K, 60 K, and 80 K. The parameters of Heisenberg Hamiltonian are the following: $J=J'=7.5$~meV, $K=0.34$~meV, $\mu=2.4\mu_\text{B}$. Magnetization curves averaged over 1000 cycles of the kMC simulations are shown with points. Solid lines correspond to the solutions of the equation (\ref{eq40}).
}
\end{center}
\end{figure}

We will not separately discuss the formula (\ref{eq61b}) for the reversal time of the magnetization of the chain $\tau^\text{strong}_\text{B}$ in the external magnetic field. Instead, we proceed to calculation of the magnetization curves $M(B)$. To do this, we need to solve the equation (\ref{eq40}) numerically. It is obvious that the obtained magnetization curves will agree with the results of the kMC simulations only if the formula (\ref{eq61b}) gives a correct estimate of $\tau_\text{B}$ in any external magnetic field $B$. Following the work~\cite{NJPhys11.063004} we start from a strong field $B_0=-5$~T. The field strength increases by an increment 0.001~T gradually to 5 T. Then the field decreases back to $B_0$, so that a sweeping cycle is complete. We consider that the magnitude of the sweeping rate of the external magnetic field $|dB/dt|$ is 130 T/s. The results of the kMC simulations are averaged over 1000 cycles.

Figure~\ref{fig15} shows the magnetization curves of the biatomic Co chain of type A at three different temperatures: 40 K, 60 K and 80 K. If temperature increases from 40 K to 60 K then the coercive field $B_C$ of the chain drops almost to zero, but the chain remains ferromagnetic. It is clearly seen from the almost constant slope of the hysteresis loop obtained by means of the kMC method. If the temperature increases to 80~K then the angle of the slope decreases significantly, which corresponds to the transition of the chain to the paramagnetic state. Thus, we can roughly estimate the critical temperature of the biatomic chain as $T_\text{C}=70\pm10$~K~\footnote{The exact determination of the critical temperatures is beyond the scope of this article. Here we only want to demonstrate the adequacy of our model within its applicability limits.}. The solid curves in Figure~\ref{fig15} show the magnetization curves obtained by solving the equation (\ref{eq40}). We see that the agreement with the results of the kMC simulations is very good at the temperature of 40 K. But the magnetization curve is slightly different from the results of the kMC simulations at the temperature of 60 K. In our opinion, such agreement is quite satisfactory. The magnetization curves obtained in the single-domain approximation become more narrow, but do not change their slope with increasing temperature. A single-domain approximation gives inadequate results if the temperature approaches to the critical one.

\begin{figure}[htb]
\begin{center}
\includegraphics[width=0.95\linewidth]{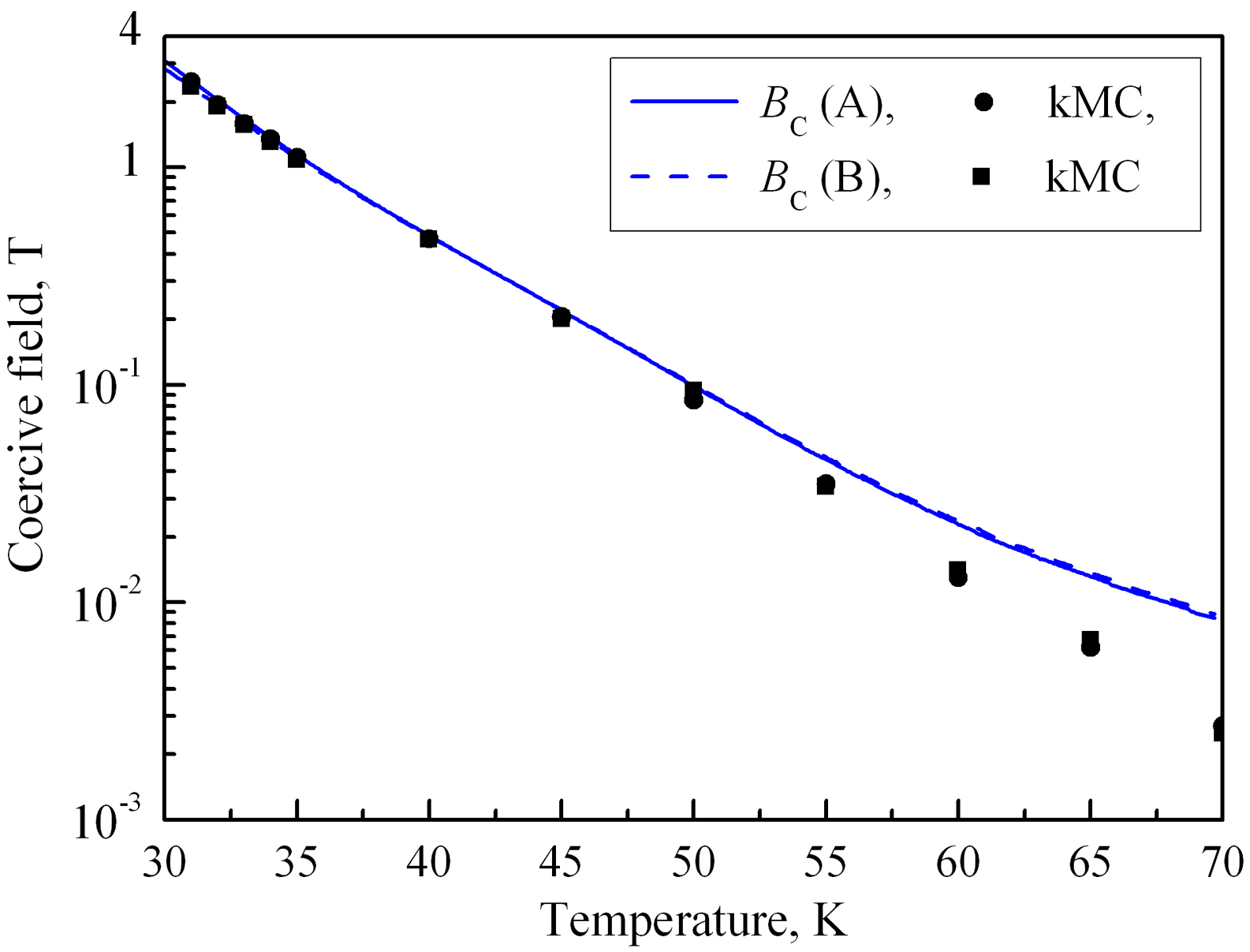}
\caption{\label{fig16} Temperature dependence of the coercive field $B_C$ for the biatomic chains of $2N=200$ atoms. The parameters of Heisenberg Hamiltonian are the following: $J=J'=7.5$~meV, $K=0.34$~meV, $\mu=2.4\mu_\text{B}$.
Results of the kMC simulations averaged over 1000 cycles are shown with points.
Solid and dashed lines correspond to the solutions of the equation (\ref{eq40}) for the chains of types A and B, respectively.
}
\end{center}
\end{figure}

Figure~\ref{fig16} shows the temperature dependence of the coercive field of the biatomic chain at $T\le T_\text{C}$. The coercive field of the biatomic chains of types A and B slightly differ from each other (the difference is less than 10\%) because the chains are quite long ($2N=200$). A single-domain approximation leads to the overestimation of the coercive fields: less than 5\% at the temperatures of $T\le40$~K, 16\% at 50 K, 76\% at 60 K, more than twice at $T\ge65$~K. If the agreement within 20\% is considered as satisfactory, then we can conclude that a single-domain approximation agrees well with the results of the kMC simulations at $T<T_\text{max}\approx0.7T_\text{C}$. This conclusion agrees with the estimate of $T_\text{max}$ obtained using the formula (\ref{eq31})~\cite{Kolesnikov1}.

\begin{figure}[htb]
\begin{center}
\includegraphics[width=0.95\linewidth]{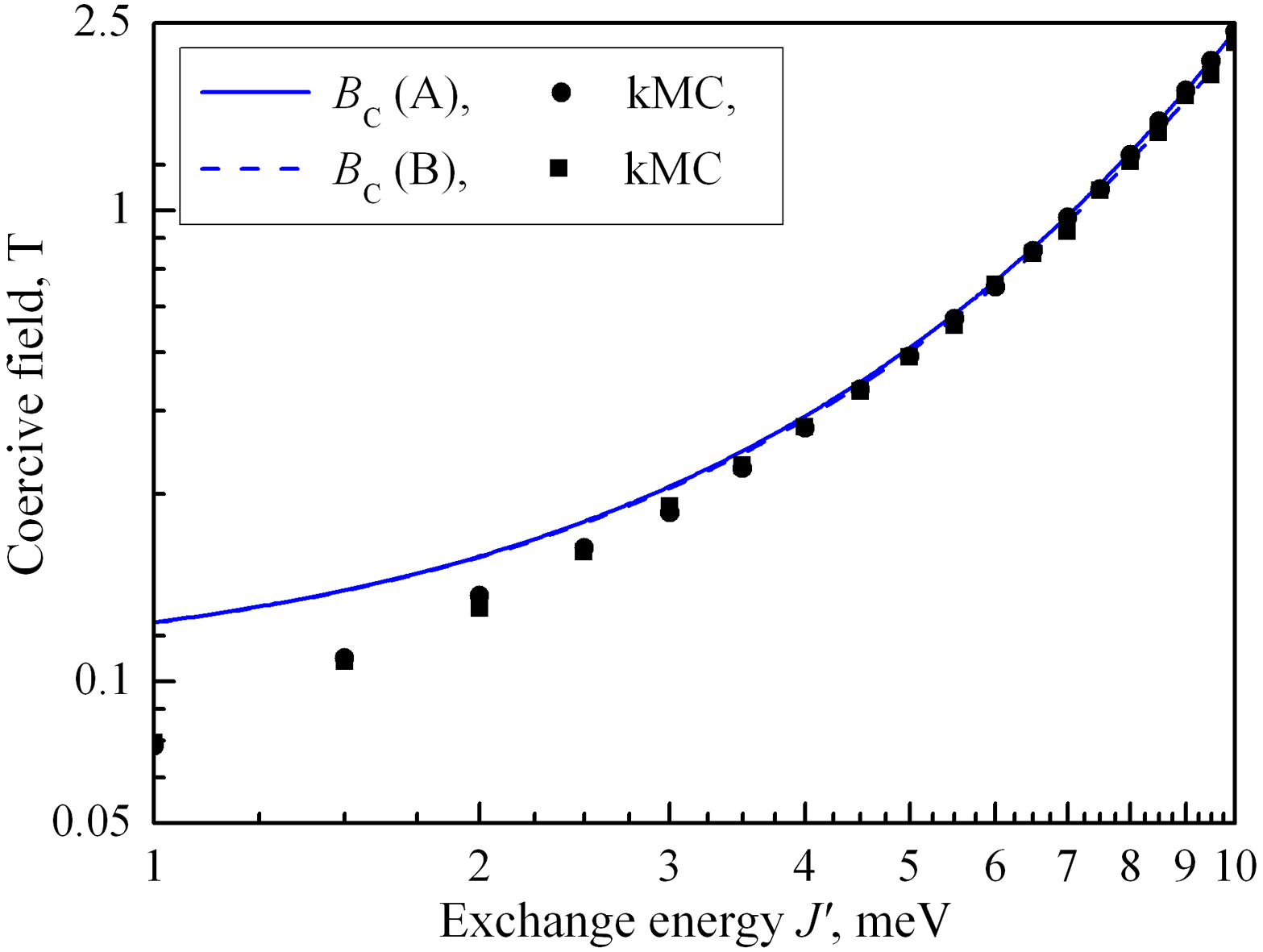}
\caption{\label{fig17} Dependencies of the coercive field $B_C$ on the exchange energy $J'\in[1,10]$~meV. The other parameters of Heisenberg Hamiltonian are the following: $J=7.5$~meV, $K=0.34$~meV, $\mu=2.4\mu_\text{B}$, $2N=200$. Results of the kMC simulations averaged over 1000 cycles are shown with points. Solid and dashed lines correspond to the solutions of the equation (\ref{eq40}) for chains of types A and B, respectively.
}
\end{center}
\end{figure}

Let us discuss the applicability limits of a strong coupling approximation. Figure~\ref{fig17} shows the dependence of the coercive field on the exchange energy $J'$. The condition of applicability of a strong coupling approximation remains the same as in the case of $B=0$ ($J'\gtrsim J/\ln N$) because  $\mu_\text{B}|B_0|=0.29$~meV and $\mu_\text{B}|B_0|\ll J,J'$. We see that the coercive field obtained in the framework of a strong coupling approximation differs from the results of the kMC simulations  by less than 5\% at $J'\ge4.5\mbox{~meV}\approx2.5J'/\ln N$. If $J'=2\mbox{~meV}\approx J'/\ln N$ then the results differ by about 20\%. Thus, we can conclude that a strong coupling approximation can be used to obtain qualitative results at $J'\gtrsim J/\ln N$, as well as in the absence of an external magnetic field. Note that the estimate obtained in a strong coupling approximation is the upper limit of the value of the coercive field at any $J'$.

The MAE of Co atoms on Pt(997) surface varies from 0.13~meV/atom for a monolayer to 2.0~meV for a single adatom~\cite{Gambardella.Nature}. We found that the coercive field of the biatomic chain remains almost constant (1.14~T and 1.11~T for the chains of types A and B, respectively) if the MAE varies in this range. These results are in a good agreement with the kMC simulations. Finally, the size effect in the range of $N\in[60,200]$ atoms was investigated. We found that the coercive field increases by about 2\% with an increase in the length of the biatomic chain in this range. These results are also in a good agreement with the results of the kMC simulations.

\section{Conclusion}\label{conc}

In summary, we considered the remagnetization of the biatomic chains in the framework of Heisenberg model with uniaxial magnetic anisotropy and a single domain-wall approximation. In general case the calculation of the reversal time of the magnetization of the biatomic chains is quite a difficult task. Therefore, we considered two limiting cases: a weak and a strong coupling between the atomic chains. In the both cases, the problem of the remagnetization of the biatomic chains is reduced to the problem of the remagnetization of the single-atomic chains. The formulas for estimating the reversal times of the magnetization of the biatomic chains in three different cases are derived: (i) the spontaneous remagnetization, (ii) the remagnetization under the interaction with the STM tip, and (iii) the remagnetization in the external magnetic field parallel to the easy axis of magnetization. The first two cases relate to both ferromagnetic and antiferromagnetic chains. The third case relates to ferromagnetic chains. For these chains, we also developed a method for calculation of magnetization curves and the coercive field.

Let us summarize the applicability limits of our method. A single domain-wall approximation is valid in a wide range of temperatures from the very low quantum tunneling temperature $T_\text{QT}$ to the maximal temperature $T_\text{max}<T_\text{C}$. The numerical estimates show that in practically important cases the maximal temperature $ T_\text{max}$ is higher than $0.7T_\text{C}$. In order to eliminate the superparamagnetic regime, the following conditions must be satisfied: $K(2N)-2(J+J')\gtrsim k_\text{B}T$ for the chains of type A and $K(2N)-2\max(J,J')\gtrsim k_\text{B}T$ for the chains of type B. The approximation of a weak coupling between the atomic chains is valid if $J'N\lesssim J$, and a strong coupling approximation is valid if $J'\ln N\gtrsim J$. In the middle range $J'\ln N\lesssim J\lesssim J'N$, both of the approximations do not give quantitative agreement with the results of the kMC simulation. However, the function $\min(\tau^\text{weak}_{(\text{STM})},\tau^\text{strong}_{(\text{STM})})$ can be used in this range, both to estimate the upper value of $\tau_{(\text{STM})}$, and for a qualitative explanation of the behavior of the $\tau_{(\text{STM})}(J')$ and $\tau_{(\text{STM})}(N)$ dependencies.

It is necessary to underline that the presented analytical method is incomparably less time-consuming than the usual kMC simulations, especially in the cases of low temperatures or long chains. Therefore, the proposed method can be a useful tool for analyzing the magnetic properties of a wide class of biatomic chains.

\section*{Acknowledgements}
The research is carried out using the equipment of the shared research facilities of HPC computing resources at Lomonosov Moscow State University~\cite{NIVC}.

\section*{Appendix. Remagnetization of single-atomic chains}

Here we summarize the main results of the previous investigations~\cite{Kolesnikov1,Kolesnikov2}. We use the same notations of the rates as in the Section~\ref{RES_AND_DISC}. Note, these notations differ from the original ones. If ${\bf B}=0$, then the reversal time of the magnetization of the ferromagnetic or antiferromagnetic single-atomic chain can be obtained as (see also Fig.~\ref{fig1})
\begin{multline}\label{eq25}
\tau=\frac{1}{2a}\left\{\frac{a}{\nu''_3}\left(\frac{N-1}{2}\right)\left[N-\frac{2(1-2a)}{1-a}\right]\right.+\\
+\left.\frac{1}{\nu''_1}\left[N(1-a)-2(1-2a)\right]\right\},
\end{multline}
where $a=\nu''_3/(\nu''_2+\nu''_3)$.

If the first atom of the chain interacts with the STM tip, then the reversal time of the magnetization is equal to
\begin{equation}\label{eq1a}
\tau_\text{STM}=\frac{1}{\nu''_3}\left(\frac{N-1}{2}\right)\left[N-\frac{2(1-2a)}{1-a}\right].
\end{equation}
The values of $\tau$ and $\tau_\text{STM}$ are related as follows
\begin{equation}\label{eq2a}
\tau=\frac{1}{2}\left\{\tau_\text{STM}+\frac{1}{a\nu''_1}\left[N(1-a)-2(1-2a)\right]\right\}.
\end{equation}
The obtained formula is applicable under the conditions: (i) $KN-2J\gtrsim k_\text{B}T$, and (ii) $T<T_{{\rm max}}$, where temperature $T_{{\rm max}}$ can be found from the equation
\begin{equation}\label{eq31}
\frac{(\nu''_1+\nu''_2+\nu''_3)(\nu''_2+\nu''_3)}{\nu''_1\nu''_2}=\left(\frac{N}{2}-1\right)^2.
\end{equation}

If ${\bf B}\ne0$, then the reversal time of the magnetization of the ferromagnetic chain ($J>0$) is equal to
\begin{multline}\label{eq38}
\tau_\text{B}(B)=\frac{1}{2(1-a_{-})}\left\{\frac{a_{-}}{\nu''_{3-}}+\right.\\
+\frac{(N-2)(1-a_{-})+(a_{-}-\alpha)S_{N-2}}{\nu''_{3+}(1-\alpha)}+\\
+\left.\frac{S_{N-2}-(a_{-}+\alpha a_{+})S_{N-3}+\alpha a_{+}a_{-}S_{N-4}}{\nu''_{1+}a_{+}}\right\},
\end{multline}
where $\alpha=(1-b)/b$, $S_N=(1-\alpha^N)/(1-\alpha)$, $a_{+}=\nu''_{3+}/(\nu''_{2-}+\nu''_{3+})$, $a_{-}=\nu''_{3-}/(\nu''_{2+}+\nu''_{3-})$, and $b=\nu''_{3+}/(\nu''_{3-}+\nu''_{3+})$.

If the MAE $K'$ and the exchange energy $J'$ of edge atoms are different from $K$ and $J$ of all other atoms, then
\begin{equation}\label{eq21a}
\tau'_\text{STM}=\left[\frac{N-5}{2\nu''_3}+\frac{b}{1-b}\left(\frac{1}{\nu'_1}+\frac{1}{\nu_2}\right)\right]
\left[N-\frac{2c}{1-a'}\right],
\end{equation}
\begin{equation}\label{eq22a}
\tau'=\frac{1}{2}\left\{\tau'_\text{STM}+
\frac{1}{a'\nu_3}\frac{b}{1-b}\left[N(1-a')-2c\right]
\right\},
\end{equation}
where $a'=\nu'_1/(\nu'_3+\nu'_1)$, $b=\nu_2/(\nu_2+\nu''_3)$, $c=3-\frac{1}{b}-2a'$.


\begin{thebibliography}{1}

\bibitem{Note1}
This function is discontinuous. Instead of (\ref {eq5}) one can use, for
  example, the expression $\nu (h_i)=\nu _0\protect \qopname \relax m{min}\left
  (1,\protect \qopname \relax o{exp}(-2h_i/k_\protect \text {B}T)\right )$.
  Then the function $\nu (h_i)$ will be continuous. In any case, it does not
  affect the applicability limits of our model, because we use the same
  functions $\nu (h_i)$ for analytical estimates and the kMC simulations.

\bibitem{Note2}
The case of a weak coupling $J'\ll J$ is much simpler. The estimations in this
  limit can be obtained by generalizing the formulas (\ref {eq5b}). The
  generalization of the formulas obtained below to the case of
  antiferromagnetic chains is also not difficult.

\bibitem{Note3}
We assume that the average time of random walk of the domain wall is much less
  than the average time of the formation of new domain wall.This condition is
  satisfied if $T<T_{\protect \text {max}}$~\cite {Kolesnikov1}.

\bibitem{Note4}
In this and the following plots, the errors of the kMC simulation results do
  not exceed the size of the marker.

\bibitem{Note5}
We also tried to describe the dependence of $\tau _{(\protect \text
  {STM})}(J')$ by the function $\left [(\tau ^\protect \text {weak}_{(\protect
  \text {STM})})^{-1}+(\tau ^\protect \text {strong}_{(\protect \text
  {STM})})^{-1}\right ]^{-1}$. Unfortunately, this function is not much better
  than the function $\protect \qopname \relax m{min}(\tau ^\protect \text
  {weak}_{(\protect \text {STM})},\tau ^\protect \text {strong}_{(\protect
  \text {STM})})$ and also does not lead to quantitative agreement with the
  results of the kMC simulations in the whole range of parameters.

\bibitem{Note6}
The exact determination of the critical temperatures is beyond the scope of
  this article. Here we only want to demonstrate the adequacy of our model
  within its applicability limits.

\end{thebibliography}


\begin{thebibliography}{10}

\bibitem{JPCM16.R603} D. Sander, J. Phys.: Condens. Matter {\bf 16}, R603 (2004).

\bibitem{JPCM22.433001} A. Enders, R. Skomski, J. Honolka, J. Phys.: Condens. Matter {\bf 22}, 433001 (2010).

\bibitem{NANO6.1} H. Wang, Y. Yu, Y. Sun, Q. Chen, NANO Brief Rep. Rev. {\bf 6}, 1 (2017).

\bibitem{Nature437.671} J.V. Barth, G. Costantini, K. Kern, Nature {\bf 437}, 671 (2005).

\bibitem{Gambardella.Nature} P. Gambardella, A. Dallmeyer, K. Maiti, M.C. Malagoli, W. Eberhardt, K. Kern, C. Carbone. Nature {\bf 416}, 301 (2002).

\bibitem{Gambardella_2003} P. Gambardella, S. Rusponi, M. Veronese, S.S. Dhesi, C. Grazioli, A. Dallmeyer, I. Cabria, R. Zeller, P.H. Dederichs, K. Kern, C. Carbone, H. Brune, Science {\bf 300}, 1130 (2003).

\bibitem{Brune.Gambardella} H. Brune, P. Gambardella, Surf. Sci. {\bf 603} 1812 (2009).

\bibitem{SSR56.189} J. Bansmann, S.H. Baker, C. Binns, J.A. Blackman, J.-P. Bucher, J. Dorantes-Davila, V. Dupuis, L. Favre, D. Kechrakos, A. Kleibert, K.-H. Meiwes-Broer, G.M. Pastor, A. Perez, O. Toulemonde, K.N. Trohidou, J. Tuaillon, Y. Xie, Surf. Sci. Rep. {\bf 56}, 189 (2005).

\bibitem{PRL102.257203} T. Balashov, T. Schuh, A.F. Tak\'acs, A. Ernst, S. Ostanin, J. Henk, I. Mertig, P. Bruno, T. Miyamachi, S. Suga, W. Wulfhekel, Phys. Rev. Lett. {\bf 102}, 257203 (2009).

\bibitem{JPCM15.R1} J. Shen, J.P. Pierce, E.W. Plummer, J. Kirshner, J. Phys.: Condens. Matter {\bf 15}, R1 (2003).

\bibitem{PRB56.2340} J. Shen, R. Skomski, M. Klaua, H. Jenniches, S.S. Manoharan, J. Kirschner, Phys. Rev. B {\bf 56}, 2340 (1997).

\bibitem{NJPhys17.023014} B. Dup\'e, J.E. Bickel,Y. Mokrousov, F. Otte, K. von Bergmann,A. Kubetzka, S. Heinze, R. Wiesendanger, New J. Phys. {\bf 17}, 023014 (2015).

\bibitem{PRL93.077203} P. Gambardella, A. Dallmeyer, K. Maiti, M.C. Malagoli, S. Rusponi, P. Ohresser, W. Eberhardt, C. Carbone, K. Kern, Phys. Rev. Lett. {\bf 93}, 077203 (2004).

\bibitem{PRB89.205426} F. Otte, P. Ferriani, S. Heinze, Phys. Rev. B {\bf 89}, 205426 (2014).

\bibitem{NJP4.100} R. F\'elix-Medina, J. Dorantes-D\'avila, G.M. Pastor, New J. Phys. {\bf 4}, 100 (2002).

\bibitem{JPCM28.503002} M. Martins, W. Wurth, J. Phys.: Condens. Matter {\bf 28}, 503002 (2016).

\bibitem{JETPL99.646} A.L. Klavsyuk, S.V. Kolesnikov, A.M. Saletsky,  JETP Lett. {\bf 99}, 646 (2014).

\bibitem{PRB70.224419} S. Pick, V.S. Stepanyuk, A.L. Klavsyuk, L. Niebergall, W. Hergert, J. Kirschner, P. Bruno, Phys. Rev. B {\bf 70}, 224419 (2004).

\bibitem{NuturePhys14.213} O. Gomonay, V. Baltz, A. Brataas, Y. Tserkovnyak, Nat. Phys. {\bf 14}, 213 (2018).

\bibitem{RevModPhys90.015005} V. Baltz, A. Manchon, M. Tsoi, T. Moriyama, T. Ono, Y. Tserkovnyak, Rev. Mod. Phys. {\bf 90}, 015005 (2018).

\bibitem{LTP40.17} E.V. Gomonay, V.M. Loktev, Low Temp. Phys. {\bf 40}, 17 (2014).

\bibitem{NatureNanotech11.231} T. Jungwirth, X. Marti, P. Wadley, J. Wunderlich, Nat. Nanotechnol. {\bf 11}, 231 (2016).

\bibitem{NatureNano10.40} S. Yan, D.-J. Choi, J.A.J. Burgess, S. Rolf-Pissarczyk, S. Loth, Nat. Nanotechnol. {\bf 10}, 40 (2015).

\bibitem{science335.196} S. Loth, S. Baumann, C.P. Lutz, D.M. Eigler, A.J. Heinrich, Science {\bf 335}, 196 (2012).

\bibitem{PRB92.184406} M. Etzkorn, C.F. Hirjibehedin, A. Lehnert, S. Ouazi, S. Rusponi, S. Stepanow, P. Gambardella, C. Tieg, P. Thakur, A.I. Lichtenstein, A.B. Shick, S. Loth, A.J. Heinrich, H. Brune, Phys. Rev. B {\bf 92}, 184406 (2015).

\bibitem{PRB92.174407} A. Ferr\'on, J.L. Lado, J. Fern\'andez-Rossier, Phys. Rev. B {\bf 92}, 174407 (2015).

\bibitem{PRB94.085406} D.-J. Choi, R. Robles, J.-P. Gauyacq, M. Ternes, S. Loth, N. Lorente, Phys. Rev. B {\bf 94}, 085406 (2016).

\bibitem{PRB86.245416} M.C. Urdaniz, M.A. Barral, A.M. Llois, Phys. Rev. B {\bf 86}, 245416 (2012).

\bibitem{PCCP17.26302} K. Tao, Q. Guo, P. Jena, D. Xue, V.S. Stepanyuk, Phys. Chem. Chem. Phys. {\bf 17}, 26302 (2015).

\bibitem{PRB90.195423} M.C. Urdaniz, M.A. Barral, A.M. Llois, A. Sa\'ul, Phys. Rev. B {\bf 90}, 195423 (2014).

\bibitem{PRL110.087201} J.-P. Gauyacq, S.M. Yaro, X. Cartoixa, N. Lorente, Phys. Rev. Lett. {\bf 110}, 087201 (2013).

\bibitem{JPCM27.455301} J.-P. Gauyacq, N. Lorente, J. Phys.: Condens. Matter {\bf 27}, 455301 (2015).

\bibitem{EPL109.57001} F. Delgado, S. Loth, M. Zielinski, J. Fern\'andez-Rossier, EPL {\bf 109}, 57001 (2015).

\bibitem{PRB93.161412R} K. Tao, O.P. Polyakov, V.S. Stepanyuk, Phys. Rev. B {\bf 93}, 161412(R) (2016).

\bibitem{RPP74.096501} H. Ebert, D. K\"odderitzsch, J. Min\'ar, Rep. Prog. Phys. {\bf 74}, 096501 (2011).

\bibitem{Landau} L. D. Landau and E. Lifshitz, Phys. Z. Sowjetunion 8, 153 (1935).

\bibitem{PhysRevB.73.174418} Y. Li, B.-G. Liu, Phys. Rev. B {\bf 73}, 174418 (2006).

\bibitem{JMMM378.186} J. Li, B.-G. Liu, J. Magn. Magn. Mater. {\bf 378} 186 (2015).

\bibitem{NJPhys11.063004} A.S, Smirnov, N.N. Negulyaev, W. Hergert, A.M. Saletsky, V.S. Stepanyuk, New J. Phys. {\bf 11}, 063004 (2009).

\bibitem{PRL96.217201} Y. Li, B.-G. Liu, Phys. Rev. Lett. {\bf 96}, 217201 (2006).

\bibitem{CPB24.097302} K.M. Tsysar, S.V. Kolesnikov, A.M. Saletsky,  Chin. Phys. B {\bf 24}, 097302 (2015).

\bibitem{PSS57.1513} S.V. Kolesnikov, K.M. Tsysar, A.M. Saletsky, Phys. Solid State {\bf 57}, 1513 (2015).

\bibitem{PhysLettA374.2058} K.-C. Zhang, B.-G. Liu, Phys. Lett. A {\bf 374}, 2058 (2010).

\bibitem{PRB93.035444} D.I. Bazhanov, O.V. Stepanyuk, O.V. Farberovich, V.S. Stepanyuk, Phys. Rev. B {\bf 93}, 035444 (2016).

\bibitem{MPLB.31.1750142} K.M. Tsysar, S.V. Kolesnikov, I.I. Sitnikov, A.M. Saletsky, Mod. Phys. Lett. B {\bf 31}, 1750142 (2017).

\bibitem{Kolesnikov1} S.V. Kolesnikov,  JETP Lett. {\bf 103}, 588 (2016).

\bibitem{Kolesnikov2} S.V. Kolesnikov, I.N. Kolesnikova, J. Exp. Theor. Phys. {\bf 125}, 644 (2017).

\bibitem{Glauber1.1703954} R.J. Glauber, J. Math. Phys. {\bf 4}, 294 (1963).

\bibitem{JCP132.134104} B. Puchala, M.L. Falk, K. Garikipati,  J. Chem. Phys. {\bf 132}, 134104 (2010).

\bibitem{PMA76.565} M. Athenes, P. Bellon, G. Martin, Philos. Mag. A {\bf 76}, 565 (1997).

\bibitem{NIVC} V. Sadovnichy, A. Tikhonravov, V. Voevodin, and V. Opanasenko, Contemporary
High Performance Computing: From Petascale toward Exascale (Boca Raton, United
States), Chapman Hall/CRC Computational Science, Boca Raton, United States,
 283307 (2013).

\end{thebibliography}
\end{document}